\documentclass[11pt, a4paper, authoryear ]{elsarticle}
\usepackage{natbib}
\usepackage{amssymb}
\usepackage{amsmath}
\usepackage{amsthm}
\usepackage{hyperref}
\usepackage[utf8]{inputenc}
\usepackage{graphicx}
\usepackage{subcaption}
\graphicspath{{./Figures/}}
\usepackage[labelformat=simple]{subcaption}

\usepackage[dvipsnames]{xcolor}
\newcommand{\RefA}[2]{{#2}}
\newcommand{\RefB}[2]{{#2}}

\journal{OMEGA, and published in https://doi.org/10.1016/j.omega.2021.102392}

\begin{document}

\begin{frontmatter}
\title{\Huge Designing optimal masks for a \\multi-object spectrometer}
\author{Juan-Jos\'e Salazar-Gonz\'alez}
\ead{jjsalaza@ull.es}
\address{Departamento de Matem\'aticas, Estad\'istica e Investigaci\'on Operativa,\\
Facultad de Ciencias, Universidad de La Laguna, Tenerife, Spain}
\date{1 January 2021}
\begin{abstract}
This paper concerns a new optimization problem arising in the management of a
multi-object spectrometer with a configurable slit unit.
The field of view of the spectrograph is divided into contiguous and parallel spatial bands,
each one associated with two opposite sliding metal bars that can be positioned to observe one astronomical object.
Thus several objects can be analyzed simultaneously within a configuration of the bars called a mask.
Due to the high demand from astronomers, pointing the spectrograph's field of view to the sky, rotating it,
and selecting the objects to conform a mask is a crucial optimization problem for the efficient use
of the spectrometer. The paper describes this problem, presents a Mixed Integer Linear Programming formulation
for the case where the rotation angle is fixed,
presents a non-convex formulation for the case where the rotation angle is unfixed,
describes a heuristic approach for the general problem, and discusses computational results on real-world and randomly-generated instances.
\end{abstract}

\end{frontmatter}

\section{Introduction}
\label{se:intro}

Is there other intelligent life in the Universe? Are there other planets where humans could live?
Scientists are constantly observing the sky looking for answers to these and other questions.
Using spectroscopy, they have detected more than 4000 planets so far, and there are many areas of the sky still waiting to be explored.
Spectroscopy splits the light from a distant celestial object into its component colours by passing it through a dispersive element (grism).
As planets orbit a star, they cause it to wobble ever so slightly.
By observing the star's spectrum, scientists can detect a slight shift in the location of the elemental absorption lines,
compared to where they should be, thus indicating that a planet was making the star wobble.
In addition, spectroscopy also informs us about the age, temperature, mass and surface gravity of objects.
The infrared spectrum provides invaluable information about the warm dust and gas phase of the Universe.
Micron-sized particles such as silicates, silicon carbide, carbon, coals, aluminum oxides or
polycyclic aromatic hydrocarbon molecules are major contributors to thermal-infrared emission.

Requests of observations submitted  to modern telescopes by scientists are carried out in queue-scheduled service mode.
Not surprisingly, spectroscopes are in high demand, and observing time is a limited resource leading to an important optimization problem in Astronomy.
Modern spectroscopes can observe dozens of objects simultaneously, but still not all desired observations can be granted. Astronomers submit proposals with objects to observe and priorities. Each proposal should be partitioned in \emph{masks} for the spectrograph. A mask is a group of objects that can fit in the same field of view of the spectrograph to be observed simultaneously, and therefore should conform to a set of technical requirements. The list of masks related to a proposal should arrive to the telescope sorted according to the sum of their objects' priorities. The masks with the highest sum of priorities are loaded and executed sequentially in a night while sky conditions and awarded time allow.
Typically around 5 to 10 masks from each awarded applicant will be completed by the instrument, so the complete partition of all the objects in the proposal (catalogue) is unnecessary.
The applicant is responsible for creating the list of masks once informed by the committee about his/her awarded time.
The necessity is for an automatic procedure to design the best-possible mask from a given catalogue. Designing a mask means selecting the center where the spectrograph must point,
selecting the rotation of the field of view, and selecting the objects satisfying the technical requirements.
It is called the \emph{Mask Design Problem} (MDP) and the aim of this manuscript is to describe an approach to solve it in practice.

\RefA{1}{The field of view of an spectrograph can be seen as a rectangle, and the MDP looks for locating the rectangle
in the celestial sphere and selecting objects inside to be observed. There are other applications where related problems need to be solved.
\RefB{4}{As an example, \cite{Radio2010} describes an optimization problem in Radiotherapy to move and configure an instrument to better deal with a tumor. Other examples of planning approaches for the efficient management of instruments are described in \cite{binter2020} 
and \cite{binter2014}.}
Real-world applications also include other side constraints (e.g. specificities of the instrument), thus the optimization problems are not identical and the approaches may not be adaptable from one to the other.
Still, we believe that the results in our paper may suggest similar approaches in other areas different from Astronomy.  }

Section \ref{se:problem} describes the context motivating the need for solving the MDP.
While the problem is described to fully capture the specificities of a modern spectrograph using a configurable slits unit, the model and algorithm
in the paper can also be adapted to traditional spectrographs where the masks are manually drilled with a precise laser cutter.
Section \ref{se:model} formulates three variants of the MDP. In the simplest variant, the center and the rotation angle
are given, so the problem is a pure combinatoric question about the objects to observe.
A more complicated variant occurs when only the rotation angle is given, so the problem also includes deciding the center for the field of view. The third variant is the MDP, and for this one we propose an approach that
enumerates some angles and applies the approach of the previous variants to each one.
Section \ref{se:results} analyses some computational experiments on real-world and randomly-generated instances
to measure the performance of the approach.

\section{Problem description}
\label{se:problem}

The research in this paper was motivated by a need at the ``Gran Telescopio de Canarias'' (GTC)\footnote{\url{http://www.gtc.iac.es/}}, which is one of the largest and most advanced telescope in the world, and one of the Spanish Unique Scientific and Technical Infrastructures (ICTS)\footnote{\url{http://www.ciencia.gob.es/portal/site/MICINN/ICTS}}.
It is a 10.4 meter class telescope located at the ``Observatorio del Roque de los Muchachos'' in the island of La Palma (Canary Islands, Spain).
The scientific production of this optimal telescope started in March of 2009 with the scientific instrument OSIRIS\footnote{\url{http://www.gtc.iac.es/instruments/osiris/}}.
The Regional Government of the Canary Islands, the Spanish Government, European Funds, and also non-European partners actively support the GTC.
Currently the telescope has three mirrors and scientific instruments to analyse the light from distant objects, and store data.
The telescope itself is enclosed by a dome, and the control of the telescope and the instruments is done remotely.
It is a very sophisticated infrastructure weighing more than 400 tons floating on hydraulic oil that can be smoothly moved with high precision and with very little force.

In 2020, GTC hosts up to four instruments (OSIRIS, MEGARA, HORuS and EMIR) which are multi-object spectrographs.
One of the most recent and permanent instruments in GTC is the ``Espectr\'ografo Multiobjeto Infra-Rojo''
(EMIR)\footnote{\url{http://www.iac.es/proyecto/emir/},\; \url{http://www.gtc.iac.es/instruments/emir/}}.
It is a wide-field near-infrared camera-spectrograph imager which is able to observe up to 55 objects simultaneously
in a spectroscopic view field of $4 \times 6.632$ square arcminutes. 
There are a few similar spectrographs today in other telescopes around the world (e.g. MOSFIRE\footnote{\url{https://www2.keck.hawaii.edu/inst/mosfire/home.html}} with a field of view of $6.12\times 6.12$ square arcminutes and able to observe up to 46 objects simultaneously in the W.M. Keck Observatory).
To understand the magnitude of these numbers in the sky, a complete rotation (turn) is 360 degrees; one degree is 60 arcminutes; one arcminute is 60 arcseconds;
the moon's angular diameter is 30 arcminutes, Venus's angular diameter is 1 arcminute,  and Proxima Centauri (the nearest-known star to the Sun) has a size of 0.001 arcsecond.

In a particular time of the year and from a particular telescope, an astronomer is not interested in observing objects dispersed within the whole astronomical sphere.
Astronomers work with astronomical catalogues of objects sharing a common type, morphology, origin, means of detection,  method of discovery and region in the sky.
Typical catalogues for an astronomer is a collection of points in an area of the sky of about one square degree (i.e. $60\times 60$ square arcminutes),
in some cases extracted from astronomical surveys like UKIDSS described in \cite{UKIDSS}.

\begin{figure}
\centering\includegraphics[width=7cm,keepaspectratio]{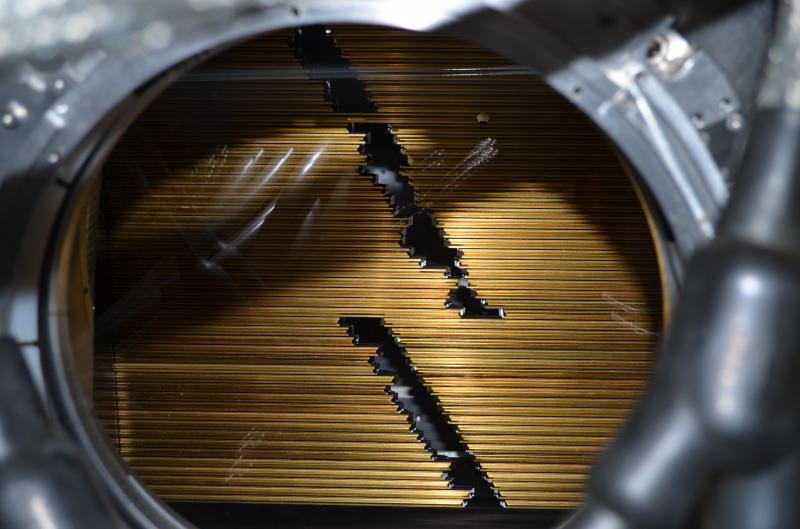}
\caption{CSU of EMIR and its fifty-five pairs of metal bars}
\label{fi:emir0}
\end{figure}

Historically spectrographs attached to telescopes gathered data through just a single slit, allowing spectral data
to be obtained from just a single object, and ignoring the light from other objects in the same field.
In the 1990s several techniques were developed to eliminate this waste of valuable telescope time, and
allowing spectrographs to analyze many objects simultaneously.
Recent developments in micro-machining have made it possible to create a reconfigurable multi-object spectrograph.
They include a cryogenic robotic unit that permits the remote configuration of a multi-slit pattern in its field of view,
where a mirror of the telescope forms the image of the sky. This device is called ``Configurable Slit Unit'' (CSU)\footnote{\url{http://research.iac.es/proyecto/emir/pages/technical-description/mechanics/csu.php}}
and is composed by pairs of opposite sliding metal bars that can be moved in one direction
to create a multi-slit pattern called a \emph{mask}.
Thus, the field of view \RefB{13}{(a rectangle) is divided into horizontal bands,}
each one associated with two retractable opposite metal bars.
\RefB{3}{Each band can observe at most one object in the sky by moving the left and right bars to create a slitlet around that object, and then analyze the light that comes through the slitlet during the exposure time of the mask in the telescope. Once the mask is loaded in the telescope and the pair of bars of a band has been configured for the object in that band, then the detector starts analysing the light and the metal bars cannot be moved.
For that reason, at most one object in a band can be observed with a single mask.}
As an example, the field of view of EMIR is divided into 55 bands, each one of height $h=7.235$ arcseconds.
\RefB{13}{Figure \ref{fi:emir0}  shows a
picture\footnote{\url{http://research.iac.es/proyecto/emir/pages/emir-gallery/csu.php}}
of the CSU in EMIR, which can observe at most 55 objects at the same time.
Hoever, for details that are given later in this section, a good mask for EMIR allows observing about 30 objects at the same time.}

When the metal bars are removed, the spectrograph becomes a wide-field imager, which gives a different use to the instrument;
in this paper we are concerned with the case where the bars create slits to observe several astronomical objects, known as
Multi Object Slit (MOS) mode.
This configurable device has obvious functional and operational advantages with respect to the more classical
approach of interchangeable multi-slit masks used in traditional spectroscopes.
Cutting masks is generally a high-precision job requiring several hours of work by a skilled technician using a laser cutter on brass sheets, and is unnecessary when using an spectrograph with a CSU.
The data in modern spectrographs are recorded digitally and processed automatically.
OSIRIS does not have a CSU, so it requires the design and construction of physical masks;
EMIR instead has a CSU, so the bars can be configured to allow the observation of several objects.

\begin{figure}
\centering
\unitlength 0.1mm
\linethickness{0.4pt}
\begin{picture}(400,400)(100,100)
\put(70,100){\line(1,0){450}}
\put(70,200){\line(1,0){450}}
\put(70,300){\line(1,0){450}}
\put(70,400){\line(1,0){450}}
\multiput(100,120)(10,0){45}{\line(1,0){5}}
\multiput(100,180)(10,0){45}{\line(1,0){5}}
\multiput(100,220)(10,0){45}{\line(1,0){5}}
\multiput(100,280)(10,0){45}{\line(1,0){5}}
\multiput(100,320)(10,0){45}{\line(1,0){5}}
\multiput(100,380)(10,0){45}{\line(1,0){5}}
\put(100,150){\makebox(0,0)[cc]{\tiny Band 1}}
\put(100,250){\makebox(0,0)[cc]{\tiny Band 2}}
\put(100,350){\makebox(0,0)[cc]{\tiny Band 3}}
\put(200,240){\circle*{4}}
\put(210,240){\makebox(0,0)[cc]{\tiny 1}}
\put(250,210){\circle*{4}}
\put(260,210){\makebox(0,0)[cc]{\tiny 2}}
\put(300,190){\circle*{4}}
\put(310,190){\makebox(0,0)[cc]{\tiny 3}}
\put(350,130){\circle*{4}}
\put(360,130){\makebox(0,0)[cc]{\tiny 4}}
\put(400,110){\circle*{4}}
\put(410,110){\makebox(0,0)[cc]{\tiny 5}}
\put(450,290){\circle*{4}}
\put(460,290){\makebox(0,0)[cc]{\tiny 6}}
\put(500,390){\circle*{4}}
\put(510,390){\makebox(0,0)[cc]{\tiny 7}}
\put(600,150){\makebox(0,0)[cc]{\tiny Region 1}}
\put(600,200){\makebox(0,0)[cc]{\tiny Region 2}}
\put(600,250){\makebox(0,0)[cc]{\tiny Region 3}}
\put(600,300){\makebox(0,0)[cc]{\tiny Region 4}}
\put(600,350){\makebox(0,0)[cc]{\tiny Region 5}}
\end{picture}
\caption{Assignment of bars to observations.}\label{fi:emir}
\end{figure}
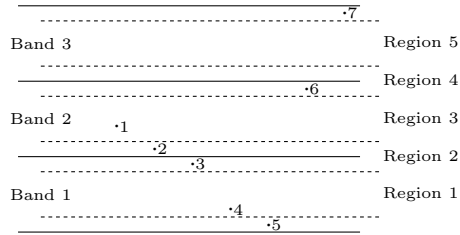

Each band of a CSU consists of three virtual horizontal zones as illustrated in Figure \ref{fi:emir}, representing
a CSU with three bands and seven objects in the field of view.
Adjacent bands in the figure are separated by continuous lines, and adjacent zones in a band by discontinuous lines.
When an object is placed in the central zone, its light will be observed by the pair of bars of that band.
When an object is placed in an upper or lower zone, then it will need two pairs of bars:
the bars of the band where the object is, and the bars of the next adjacent band.
When an object is placed exactly on the border between the central zone and an upper (or lower) zone,
it is considered in the central zone, so only the bars of that band are required for observing this object.
Thus, object 1 in Figure \ref{fi:emir} will only need the bars in band 2 to be observed;
object 2 will need the bars of bands 1 and 2, exactly as object 3; object 4 will need the bars of band 1 only;
objects 5 and 7 cannot be observed in that field of view if bands 1 and 3 are the lowest and highest of the CSU; object 6 will need the bars of bands 2 and 3. No pair of objects in $\{1,2,3,6\}$ could be observed simultaneously because they all need the bars of band 2, thus at most one of them could be selected for the mask.
In EMIR there are 55 bands, each one with a height of 398/55 arcseconds, and the height of the upper and lower zone in a band is 1 arcsecond (thus the height of the central zone is 5.235 arcseconds).

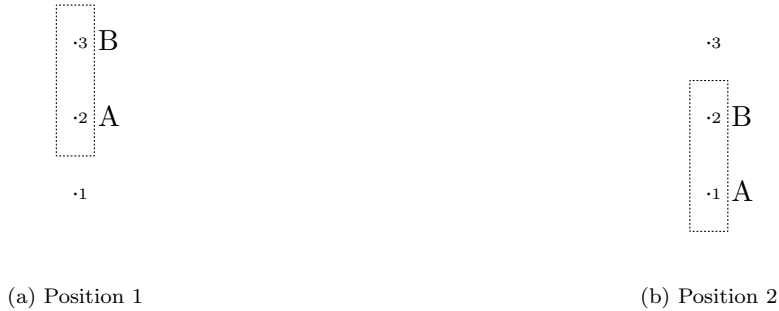
\begin{figure}

\begin{subfigure}{.5\textwidth}
\centering
\unitlength 0.1mm
\linethickness{0.4pt}
\begin{picture}(100,300)(50,0)
\put(100,100){\circle*{4}}
\put(110,100){\makebox(0,0)[cc]{\tiny 1}}
\put(100,200){\circle*{4}}
\put(110,200){\makebox(0,0)[cc]{\tiny 2}}
\put(100,300){\circle*{4}}
\put(110,300){\makebox(0,0)[cc]{\tiny 3}}
\put( 75,150){\dashbox{2.5}(50,200)[tl]{}}
\put(130,190){A}
\put(130,290){B}
\end{picture}
\caption{Position 1}\label{fi:nod1}
\end{subfigure}
\begin{subfigure}{.5\textwidth}
\centering
\unitlength 0.1mm
\linethickness{0.4pt}
\begin{picture}(100,300)(50,0)
\put(100,100){\circle*{4}}
\put(110,100){\makebox(0,0)[cc]{\tiny 1}}
\put(100,200){\circle*{4}}
\put(110,200){\makebox(0,0)[cc]{\tiny 2}}
\put(100,300){\circle*{4}}
\put(110,300){\makebox(0,0)[cc]{\tiny 3}}
\put( 75, 50){\dashbox{2.5}(50,200)[tl]{}}
\put(130, 90){A}
\put(130,190){B}
\end{picture}
\caption{Position 2}\label{fi:nod2}
\end{subfigure}
\caption{Two sky positions with the chopping-and-nodding technique}\label{fi:nod}
\end{figure}

The observation of some objects may need the bars of more than one band, even if they fall in the central zone.
The reason is that the atmosphere does not only absorb infrared photons coming from astrophysical targets, but also
emits a strong background with the spectral shape of a black body.
The sky brightness changes on short timescales and over small distances in the sky.
The telescope gives an additional background.
Consequently, the number of photons reaching the detector of the spectrograph is huge.
Because of the very high background from the ambient atmosphere and telescope, the sensitivity of ground-based infrared instruments cannot compete with that of space-borne ones. However,
ground based instruments mounted on large telescopes offer superior spatial resolution.
The basic idea to suppress the thermal background is to perform differential observations, using
a two-position technique called \emph{beam switch} or \emph{chopping-and-nodding}.
In this technique the secondary mirror of the instrument moves back and forth between two nearby sky positions.
One sky position includes two observations: the background and the astronomical object (the on-source observation or target position), say observation A, and another sky point with pure background (off-source observation), say observation B.
The other sky position has the same astronomical object as observation B, and another pure background (off-source observation) as observation A.
Figure \ref{fi:nod} illustrates the two sky positions for the telescope, each one with the two observations.
Analysing the light of an object requires observing three points in the sky: the on-source observation (i.e. the object), represented by the point with label 2, and two
 off-source observations (i.e. pure background), represented by the points with labels 1 and 3.
To observe that object the detector needs an slitlet (rectangle) that will be alternating between position 1 and position 2,
moving the telescope up and down. At each position, the detector registers the light from two points,
represented in the figure by A and B. The on-source is always under observation, either at point A when the telescope is in position 1, or at point B when the telescope is in position 2. The off-source observation is the sky point 3 when the telescope is in position 1, and the sky point 1 when the telescope is in position 2.
This observing strategy is called ``nodding on the slit" since the telescope is constantly repointing up and down.
The two positions have to be alternated at a rate faster than the rate of the background fluctuations.
In the case of point sources, the chopping and nodding throw is usually set around $t=7$ arcseconds to ensure
proper separation of the different beams.
\RefB{9}{Clearly, if  $t$ is larger than the height $h$ of a band, the on-source and off-source points will need different pairs
of bars to be observed.}
The angle $t$ is decided by the astronomer and must be the same for all the objects needing the chopping-and-nodding technique.
Instrument overheads due to chopping-and-nodding duty cycle losses are negligible,
and the observing time is divided between the two positions.
Spectroscopic acquisition time for a mask depends on the source brightness, typically between 10 and 15 minutes
which, together with the \RefB{12}{observation time,}  means that a mask will take typically one hour, or more, of use of the spectrograph.
The telescope also consumes time to point the instrument to a given location with a given rotation,
and move the bars to perform the selected observations. However, this setup time (less than one minute) is negligible compared to the observation time for the selected objects.

A fundamental requirement to properly analyze the light coming from a particular point without disturbance by the light of another object is that no object can be at a distance to that point smaller than a given threshold (3 arcseconds in practice), except the object associated with that point. Imposing this requirement to on-source points is easy in a preprocessing phase, as it forces removing from the catalogue all objects closer than the threshold to other objects in the sky. Therefore, the input to the MDP is not only the set of objects of interest to the astronomer,
but also all objects in the sky that may affect the observation of an object of interest.
Imposing the requirement to off-source points is also easy once the rotation angle for the mask has been decided.
Indeed, we must remove from the list of objects of interest the ones for which there exists another object in the sky too close to an off-source point. The major difference between the two procedures is that an on-source removal is permanent while an off-source removal is temporal, valid only while looking for a mask with that particular rotation angle.

While the instrument does this chopping-and-nodding movement to observe all the required points in the mask (alternating positions every $t$ seconds, with this parameter decided by the astronomer), objects with brightness larger than a given threshold do not need their off-source observations and the on-source points are observed half of the awarded time.
This strategy makes some bands available to potentially observe other objects.
\RefB{5}{When there are two brilliant objects $i$ and $j$ located in the sky points 1 and 2, respectively, at Figure \ref{fi:nod}, one could think that the band observing $j$ when the telescope is in position 1 could also observe $i$ when the telescope is in position 2. However, in practice, this assignment of two celestial objects to the same band in a mask is technically forbidden because each pair of bars must collect and analyze the light of (at most) one object.
The approach in this manuscript keeps this requirement, although it can be adapted easily to the case where two brilliant objects could be assigned to the same band.
Another technical requirement is that brilliant objects (i.e. not needing the off-source observation) should lie on central zones, so each selected brilliant object books only one band in the mask. The approach in this manuscript keeps this other requirement as well, although it can be adapted easily to the case where the requirement is removed.}

Each object in the catalogue is marked to know whether the pair of bars must analyze either the three chopping-and-nodding points or only the on-source point, if the object is selected to be observed. Each object has also a priority (for example depending on its position in a Colour Magnitude Diagram\footnote{\url{https://www.schoolsobservatory.org/discover/projects/clusters/cmd}}).
Some of objects, in addition, are required to be located within a maximum distance with respect to the center of the CSU, if they are selected to be observed.
This restriction depends on the lines of the object's spectrum of interest to the astronomer.
In MOS mode, the central wavelength of the scattering element falls in the centrum of the band, and from there it is scattered to both sides. For that reason, depending on the spectrum of interest, the object position in the band is horizontally limited to guarantee that the spectrum image falls inside the detector.
For example, the width of a band in EMIR is four arcminutes (i.e. 240 arcseconds), and some objects may be required to be located at most one arcminute to the left from the center of the CSU and zero to the right in order to capture
a particular wavelength in the light from that object.

It is worth mentioning that,  when the chopping-and-nodding throw is quite large, the rectangle in Figure \ref{fi:nod} alternating between position 1 and position 2 could be the union of the slitlet of two non-adjacent bands. This occurs when  the chopping-and-nodding throw is, for example, larger than double the height of a band. In that case, if the object requires the chopping-and-nodding technique, the rectangle in the figure must be replaced by two disjoint rectangles (A and B), each one representing the slitlet of one or two bands required for the observation of a point. In position 1, the rectangles A and B read the points 2 and 3, respectively, and in position 2 the rectangles A and B read the points 1 and 2, respectively. The height of the rectangles is a multiple of the band height, and the \RefB{14}{width} depends of the object properties (typically around five arcseconds). There may exists objects inside a rectangle different than the object in position 2, but not at a distance smaller than a given threshold to the point under observation in that rectangle.
This threshold is decided by the astronomer, and it is typically about two arcseconds.

On a regular basis, normally each semester, astronomers around the world submit their observing proposals to a committee at the telescope, and this committee distributes the available observing time of each instrument according to the scientific merits of the applicants.
A proposal is a catalogue of (primary and secondary) objects that an astronomer wishes to observe, each one with a priority.
Due to the time limitation awarded by the  committee, not all the objects in a proposal \RefB{15}{may finally be} observed.
To specify the object selection problem, the telescope requires each applicant to build a list of masks for the spectrograph
maximizing the sum of the priorities of the observations in each mask (total priority of the mask). The masks in the list will be processed in descending order of
total priority while there is awarded time of the spectrograph for the astronomer. While in theory a complete partition of the catalogue into masks is required, in practice only between five and ten masks will be loaded and performed by the instrument.
For that reason we face the MDP, i.e. the problem of selecting the objects for one mask with maximum total priority.
Once an instance is solved, the selected objects are removed from the catalogue, and a new instance will be solved.
The precise number of required masks will depend on the observation time of the objects,
the time awarded to the astronomer by the committee, and the sky condition of the night.

\bigskip

MDP is related to locating a rectangle on a plane.
\RefB{3}{This is a research area in the Computational Geometry community with many problem variants.
There are several articles in the literature dealing with enclosing subsets of points with all kinds of geometric elements.}
Given a finite planar point set, the \emph{enclosing problem} is to find the smallest geometrical element of a given type and arbitrary orientation that encloses all the $n$ points. A kind of dual variant of the enclosing problem is finding the translation and orientation for a geometrical element of a given size to maximize the number of enclosed points.
\cite{katz2002} presents an $O(n \log n)$ time algorithm for locating an axis-parallel rectangle of given width and height.
The algorithm is based on the segment tree data structure.
\cite{Cabello2008} describes a $O(n^3  \log n)$ time algorithm to compute the translation and rotation of a unit square.
\cite{kaplan2019} presents a $O(n^{5/2} \log n)$ time algorithm to compute an axis-parallel rectangle of a given perimeter,
area or diagonal (so the length of one edge uniquely determines the length of the other edge).
Although these problems are polynomial-time solvable, their algorithms are obviously not efficient in practice when $n$ is a large number.
\RefB{3}{Having observed that locating a rectangle can be seen as a relaxation of the MDP,
it is hopeless to expect finding reasonable solutions for the MDP with an algorithm for locating a rectangle.
The reason is that there are many other side constraints in the MDP due to specificities of the application in Astronomy.}

\bigskip

While the MDP is of high interest for telescopes around the world, all previous attempts to solve it are manual.
Knowing the complexity of the optimization problem, some telescopes offer graphical user interfaces to help astronomers to solve it, but without any algorithm to automatically generate a solution, neither optimal nor near-optimal.
An example of one of these interfaces is the software called ``Optimized Slits Positioner''\footnote{\url{http://www.iac.es/proyecto/emir/pages/observing-with-emir/observing-tools/osp.php}}.
It is a guided user interface that can overlay the field of view of EMIR with the
CSU in an astronomical image.
Another example is the software called ``MOSFIRE Automatic GUI-based Mask Application''\footnote{\url{https://www2.keck.hawaii.edu/inst/mosfire/magma.html}} for the MOSFIRE field of view.
The astronomer must locate manually the instrument pointing over the image,
rotate the CSU to the best position angle,
and set a slit pattern for the bars on top of the desired  objects in the view field.
While the interface offers some capabilities to automate the slit pattern for each pair of bars once an object is given,
the other actions depend solely on the astronomer's ability.
Our paper proposes the first automatic approach to solve the MDP.

\section{Mathematical Formulation}
\label{se:model}

MDP is the problem of finding the location in the sky where the center of the field of view should point, the angle of rotation with respect to the horizon (observer's ground plane), and the astronomical objects from the catalogue to observe simultaneously. The aim is to maximize the sum of the priorities of the selected objects.
Let $(x,y)$ be the coordinates of the center, $\theta$ be the rotation angle of the field, and
$z_i$ be 0 or 1 to indicate whether an object $i$ is in the selection. These three arrays define a solution to the MDP.
\RefA{2}{This section states a general notation in Subsection \ref{se:not} and then describes three problems.
Subsection \ref{se:fix} concerns the simplest problem of finding $z_i$ for each object $i$ when $(\theta,x,y)$ are fixed to given values.
Subsection \ref{se:center} extends the previous problem to find $(x,y)$ and $z_i$ when $\theta$ is given.
Finally, Subsection \ref{se:angle} goes further with the extension, and it describes the optimization problem for finding $(\theta,x,y)$ and $z_i$, which is the MDP. A mathematical model is presented for each problem.
The models for the first and second problems are Mixed Integer Linear Programming formulations.
The model for the MDP is a non-convex formulation, difficult to be solved with today state-of-the-art solvers.
Section \ref{se:algorithm} describes a heuristic approach for solving MDP based on the formulations for the first and second
problems.}

\subsection{Notation}
\label{se:not}

Let us consider a catalogue $I=\{1,\ldots,n\}$ of celestial objects, each one $i\in I$ associated with coordinates $(a_i,b_i)$ in the sky, and a profit $p_i$ if observed.
The profit of an object is defined by the priority that such object has for the astronomer, and we assume that $I$ only contain object of interest, i.e. $p_i>0$ for each $i\in I$.
We use the horizontal coordinate system (based on altitude and azimuth) in degrees, arcminutes and arcseconds.
Thus, objects represented in equatorial coordinates (based on the right ascension and declination) or in ecliptic coordinates (based on the latitude and longitude) should be represented in horizontal coordinates.
Since catalogues in practice contain objects in a very small area (of about one square degree), horizontal coordinates can be managed as Cartesian coordinates in a 2-dimensional square tangent to the celestial sphere at the central point of the area. The approach in this article could also be adapted to work with objects directly represented in 3-dimensional spherical coordinates, which is not of interest in practice.

To simplify the mathematical constraints for the selected objects in the field of view, it is convenient to
represent each object in a coordinate system with axes parallel to the border of the field of view,
which means a rotation of angle $\theta$. Since the horizontal coordinate system may come with large numbers in one dimension and small numbers in the other dimension, with the catalogue embedded in a small area of the sky, it is also convenient for numerical purposes to translate the origin coordinate (0,0) to the centroid $(a_0,b_0)$ of the catalogue.
Applying the translation first and the rotation after, an object with horizontal coordinates $(a,b)$ will have the relative coordinates
\begin{equation}
(a',b') = \big( \, (a-a_0)\cos\theta-(b-b_0)\sin\theta \;, \;(a-a_0)\sin\theta+(b-b_0)\cos\theta \,\big). \label{eq:rot}
\end{equation}

The objects of interest are partitioned by the astronomer into two sets depending whether an object is brilliant
enough and does not need the chopping-and-nodding technique (i.e. only on-source observation),
or it has a low brightness and therefore two off-source observations at distance $t$ arcseconds are necessary in addition
to the on-source observation. Due to the design of the CSU, the off-source observations must be ``on the slit'',
meaning that they must be two points on the sky at $t$ arcseconds from the on-source observation, on the
vertical axis of the CSU, one point above the object and the other point below the object. Typically the throw $t$ is a value between 5 and 10, and it must be the same for all the objects
needing the chopping-and-nodding technique since the telescope (not the mask) will be moving up-and-down permanently. Because the two off-source points will not be observed simultaneously,
we can concentrate the design of the mask considering only one off-source point, for example the one above the object.
We use the notation $I^1$ for the brilliant objects (needing only the on-source observation) and
$I^2$ for the other objects (needing the on-source and the off-source observations separated by $t$ arcseconds).

We assume that $I$ does not contain an object too close to another object in the sky within a given threshold
(3 arcseconds in practice). This assumption guarantees that an on-source observation will not be disturbed by
the light from another nearby object.

Let us consider a spectrograph with a field of view being a rectangle with height $H$ and width $W$.
When the spectrograph has a CSU, the height of the field of view is divided into $m$ bands represented by $J=\{1,\ldots,m\}$. Each band contains two opposite bars that can create a slit at any position in the interval $[-W/2,W/2]$ to observe one astronomical object.
When no object is observed by the bars of a band, the bars are touching each other in the middle of the band (position 0).
When one object is observed by the bars of two adjacent bands (e.g. object 2 in Figure \ref{fi:emir}), the bars must create a slitlet at the same position of the object.
Each object $i$ is required to be horizontally located in a subinterval $[\underline W_i,\overline W_i]$ of the band $[-W/2,W/2]$ depending on the particular wavelength of interest to the astronomer for that object.


The width $w_i$ of the slitlet created by the pair of bars is selectable by the astronomer
according to the observing conditions and the required spectral resolution of the object $i$ to observe.
A typical value for $w_i$ is about $5$ arcseconds.
The height of a band is $h=H/m$. Each band is subdivided into three zones, as illustrated in Figure \ref{fi:emir}.
The height of the upper zone is $\overline h$ and the height of the lower zone is $\underline h$,
thus the height of the central zone is $h-\underline h - \overline h$.
An object located in the central zone (e.g. object 1) needs the bars of that band only to be observed.
An object located in the lower zone (e.g. object 2) needs the bars of two bands: the one where it is placed and the band below.
An object located in the upper zone (e.g. object 3) needs the bars of two bands: the one where it is placed and the band above.
For that reason, an object in the lower zone of the lowest band (e.g. object 3) or in the upper zone of the highest band $m$ (e.g. object 7) cannot be
selected to be observed with the current field of view.
For notational convenience we merge the upper zone of a band and the lower zone of the adjacent band into a single region with height $\underline h + \overline h$. The central zones are considered odd regions and the merged zones are considered
even regions (see Figure \ref{fi:emir}).
Let $K=\{1,\ldots,2m-1\}$ represent the set of all regions, with region 1 being the central zone of the lowest band (band 1), and region $2m-1$ being the central zone of the highest band (band $m$).
Let $l_k$ be the position in the vertical axes of the CSU where region $k$ ends respect to the center of the field of view, i.e.
\[ l_k = l_{k-1}+\begin{cases} h-\overline h -\underline h & \mbox{ if $k$ is odd} \\
\overline h + \underline h & \mbox{ if $k$ is even} \end{cases} \]
with  $l_0=\underline h-H/2$. Thus, $l_{2m-1}=H/2-\overline h$.
Let $K_j$ be the regions where objects need the bars of band $j$ to be observed.
Note that $|K_1|=|K_m|=2$ and $|K_j|=3$ for all $1<j<m$.

\subsection{Deciding the objects when the center and rotation are given}
\label{se:fix}

This section describes an exact approach to select the objects for a mask with a given center and rotation angle.
This restricted MDP is called the \emph{Object Selection Problem} (OSP).
We assume that the coordinates of the objects have been rotated using (\ref{eq:rot}), including the coordinates $(x',y')$ of the given center, so the field of view is fixed and axis-parallel. 
In this case, we also remove from $I$ all objects needing the chopping-and-nodding technique with an off-source
observation too close to another object in the sky.

Observing a point $(a',b')$ with $-W/2 \leq a'-x' \leq W/2$ and $l_{k-1} < b' - y' < l_k$ for some $k\in K$ needs the bars of band $j$ if $k=2j-1$ and the bars of bands $j$ and $j+1$ if $k=2j$.
For convention, an object $i$ with $b'_i-y'=l_k$ needs only one band, which is $j+1$ if $k=2j$ or $k=2j+1$.
If an object $i\in I^1$ is selected in the OSP then the CSU must observe one point with coordinates $(a'_i,b'_i)$,
which must fall in a central zone of a band (i.e. in an odd region).
If an object $i\in I^2$ is selected in the OSP then the telescope must observe the two points $(a'_i,b'_i)$ and $(a'_i,b'_i+t)$, so this object may require the bars of up to four bands.
Strictly speaking the telescope must observe the points $(a'_i,b'_i)$, $(a'_i,b'_i+t)$ and $(a'_i,b'_i-t)$,
but the two off-source points are not simultaneously observed, as illustrated in Figure \ref{fi:nod}.
Let $J_i$ be the bands needed for observing object $i$. Enumerating the objects in
\[ I' = \left\{ i \in I \; : \; \underline W_i \leq a'_i-x' \leq \overline W_i \;,\; \underline h -H/2 \leq b'_i-y' \leq H/2-\overline h -
\big\{\begin{array}{ll} 0     & \mbox{ if }i\in I^1 \\ t & \mbox{ if }i\in I^2 \end{array} \right\}\]
allows us to define $I'_j$ as the subset of objects in $I'$ requiring the bars of the band $j$ to be observed.
Note that if $i\in I'_j$ then there is a region $k\in K_j$ such that $l_{k-1} \leq b_i' - y' \leq l_k$.


Let us introduce a binary variable $z_i$ for each $i\in I'$ to decide whether the object $i$ is selected for the optimal mask.
Then the OSP is
\begin{eqnarray}
 \max \sum_{i\in I'} p_i \cdot z_i && \label{eq:w1}\\
\sum_{i \in I'_j} z_i \leq 1 && \forall j\in J:|I'_j|\geq 2\label{eq:w2}\\
z_i\in \{0,1\}&& \forall i\in I'.\label{eq:w3}
\end{eqnarray}
In practice, it is convenient to eliminate each object $i\in I'$ when there is another object $i'$ such that $J_{i'}\subseteq J_i$ and $p_{i'} \geq p_i$.

If $t\leq  \underline h + h + \overline h$ (which occurs in the real-world application motivating our research) then the matrix defined by inequality (\ref{eq:w2}) is
an \emph{interval matrix} since each column (object $i$) has coefficient 1 in consecutive rows (the bands in $J_i$).
Due to the Consecutive Ones Property (see e.g. \cite{Conse76}), the matrix is totally unimodular and therefore one can replace the integrability by non-negativity of the variables in (\ref{eq:w3}) and still get an integer solution.

If $t>  \underline h + h + \overline h$, we can get a matrix with the square submatrix
\[\left[\begin{tabular}{ccc}
0 & 1 & 1 \\
1 & 1 & 0 \\
1 & 0 & 1
\end{tabular}\right]\]
and then the blossom inequalities
\begin{equation}
\sum_{i\in I' : |J_i \cap J'|\geq 2 } z_i \leq \frac{ |J'|-1}{2} \qquad \forall J'\subseteq J : |J'| \mbox{ is odd } \label{eq:blossom}
\end{equation}
may help to strengthen the Linear Programming (LP) relaxation of the OSP formulation.
While there is at most one inequality (\ref{eq:w2}) for each band,
the number of blossom inequalities (\ref{eq:blossom}) is exponential in $m$.
Still, given a non-negative (fractional) solution $z^*$ of (\ref{eq:w2}),
deciding whether this solution satisfies all inequalities (\ref{eq:blossom}) or finding an odd subset $J'$ violating (\ref{eq:blossom})
(known as the \emph{separation problem}) can be solved in polynomial time (e.g. \cite{oddcut}).

Let $G$ be an edge-weighted hypergraph with a vertex for each band $j\in J$ and an hyperedge for each object $i\in I'$. The hyperedge associated to an object $i$ is $J_i$.
A \emph{matching} in the hypergraph $G$ is a set of disjoint hyperedges, and represents a feasible solution of the OSP. A \emph{perfect matching} corresponds to a selection of the objects where all the bands are used.
The intersection graph of $G$ (also called the \emph{line graph} of a hypergraph) is a graph $H$ with a vertex for each object and an edge connecting two objects when they need a common band (i.e., they cannot be together in the same mask).
The maximum-weight matching problem in $G$ is equivalent to the maximum-weight independent-set problem
(also known as the maximum weight vertex packing problem) in $H$.
Since any graph is the intersection graph of an hypergraph,
the maximum-weight matching problem of an hypergraph is $\cal{NP}$-hard (and $\cal{APX}$-hard).
However, there are some particular classes of hypergraphs on which the problem is solvable in polynomial time.
The best-known class is the one of the 2-uniform hypergraphs (i.e. the graphs) in which case the algorithm in \cite{Gabow1990} solves the maximum-weight independent set in $O(mn'+m^2 \log m)$ time, where $m=|J|$ and $n'=|I'|$.
A larger class where the problem is still solvable in polynomial time includes the hypergraphs where the intersection graph is claw-free; a claw is a three-leaf tree; see e.g. \cite{Independent2017}.
Thus, if $\underline h = \overline h = 0$ then all the extreme points of the polytope defined by (\ref{eq:w2}), (\ref{eq:blossom}) and the non-negativity are integer; in other words, OSP is an easy problem to solve.
The computational complexity of the OSP when $t>  \underline h + h + \overline h$, $\underline h >0$ and $\overline h >0$ is
unknown, but in our experiments we never got a fractional non-negative solution of the linear program
(\ref{eq:w1}), (\ref{eq:w2}) and (\ref{eq:blossom}).

\subsection{Deciding the center and the objects when the rotation is given}
\label{se:center}

Again, we assume that the coordinates of the objects are all rotated by the given angle,
and an object of interest was removed when a related observation point is too close to another object in the sky.
Now the problem is to decide the center of an axis-parallel field of view and the set of objects to observe.
We call it \emph{Center-Objects Selection Problem } (COSP), and it can be formulated as follows.

Let  $(x',y')$ be a pair of continuous variables to represent the center of the field of view,
and a binary variable $z_i$ for each $i\in I$ to decide whether the object $i$ is selected for the optimal mask.
Let us consider a binary variable $v_{ik}$ assuming value 1 if and only if object $i\in I$ is selected and located in region $k\in K$ of the optimal mask. When $i\in I^2$ we also introduce a binary variable $w_{ik}$ to indicate whether the off-source point of object $i$ falls in region $k$, and
a binary variable $u_{ij}$ to indicate whether the object $i$ needs the bars of band $j$ in the mask.
Then the COSP is
\begin{eqnarray}
\max \sum_{i\in I} p_i \cdot z_i \label{eq:obj}&&\\
\underline W_i \cdot z_i - M\cdot (1-z_i) \leq a'_i - x'  \leq \overline W_i\cdot z_i+M\cdot (1-z_i)&  & \forall i\in I \label{eq:x} \\
l_{k-1} \cdot v_{ik} - M\cdot (1-v_{ik}) \leq b'_i - y' \leq l_k\cdot v_{ik}+M\cdot (1-v_{ik}) & & \forall i\in I,k\in K  \label{eq:v}\\
l_{k-1} \cdot w_{ik} - M\cdot (1-w_{ik}) \leq b'_i +t - y' \leq l_k\cdot w_{ik}+M\cdot (1-w_{ik}) & & \forall i\in I^2,k\in K  \label{eq:w}\\
\sum_{k\in K} v_{ik} = z_i &&\forall i\in I\label{eq:vz}\\
\sum_{k\in K} w_{ik} = z_i &&\forall i\in I^2\label{eq:wz}\\
\sum_{k\in K_j} v_{ik} \leq  u_{ij} &&\forall i\in I^2, j\in J\label{eq:vu}\\
\sum_{k\in K_j} w_{ik} \leq  u_{ij} &&\forall i\in I^2, j\in J\label{eq:wu}\\
\sum_{i\in I^1}\sum_{k\in K_j} v_{ik} + \sum_{i\in I^2} u_{ij} \leq 1 &&\forall j\in J\label{eq:uz}\\
v_{ik}\in\{0,1\} &&  \forall i\in I,k\in K  \\
w_{ik}\in\{0,1\} &&  \forall i\in I^2,k\in K  \\
u_{ij}\in\{0,1\} &&  \forall i\in I^2,j\in J  \\
z_i\in\{0,1\} &&  \forall i\in I\\
x',y'\in \mathbb{R}.&& \label{eq:real}
\end{eqnarray}
The objective function (\ref{eq:obj}) guarantees to select objects leading to the maximum sum of priorities.
Inequalities (\ref{eq:x}) ensure that an object can only be selected if the horizontal distance to the center is not larger than $\underline W_i$ on one side and not larger than $\overline W_i$ on the other side.
Inequalities (\ref{eq:v}) define the region of the on-source point of object $i$ if selected for the optimal mask.
Inequalities (\ref{eq:w}) define the region of the off-source point of selected objects needing the chopping-and-nodding technique.
Equations (\ref{eq:vz}) and (\ref{eq:wz}) force the observations of the points associated with selected objects.
Inequalities (\ref{eq:vu}) and (\ref{eq:wu}) consider the band required by each off-source point.
Note that the bars of each band can observe at most one object, but they can observe both the on- and off-source points if the object needs the chopping-and-nodding technique. Inequalities (\ref{eq:uz}) force that at most one object is selected in the regions $K_j$ for each band $j$.

Although we have presented the formulation with one big-$M$ value for all the inequalities, it is convenient to strengthen the LP relaxation with the smallest possible value for each inequality. In particular, we can use $M=\max\{ a'_i : i\in I\} - \min\{a'_i:i\in I\}$ for (\ref{eq:x}),
$M=\max\{ b'_i : i\in I\} - \min\{b'_i:i\in I\}$ for (\ref{eq:v}), and
$M=\max\{ b'_i : i\in I^2\} - \min\{b'_i:i\in I^2\}+t$ for (\ref{eq:w}).
When implementing this formulation, it is important to introduce an small value $\epsilon$ and replace the values $l_{k-1}$ and $l_k$ in inequalities (\ref{eq:v}) and (\ref{eq:w}) by $l_{k-1}+\epsilon$ and $l_k-\epsilon$, respectively, when $k$ is even to ensure that a point on the border of a central region is counted only in that region.
\RefB{8}{Since objects distant from each other cannot be observed in the same mask,
there exists an incompatibility relation on the $z_i$ variables.
This incompatibility relation shows that the (Maximum) Stable Set Problem is a relaxation of the MDP,
and suggests the use of valid inequalities like the clique inequalities; see e.g. \cite{ssp}.
While the exact separation of these inequalities is a difficult problem,
modern optimizers include sophisticated built-in heuristic separation procedures,
thus we do not detail them in this paper (also because they did not show to be helpful in our experiments).} 

While the integrality of some binary variables is unnecessary (e.g., Equations (\ref{eq:vz}) force
integrality of the $z_i$ variables when the $v_{ik}$ variables are integer),
we found better performances of the optimizer when keeping them in the formulation.
\RefB{10}{In addition, if $t=rh$ for a positive integer number $r$, the model can be simplified by observing that each
off-source point is always $2r$ regions over its on-source point, i.e. $w_{i,k+2r}=v_{ik}$ for all $i\in I^2$ and $k\in K$.}

\RefB{17}{We end this subsection emphasizing that $(x',y')$ are the relative coordinates for the center of the mask, rotated with the given angle $\theta$. Then, one must apply (\ref{eq:rot}) to compute the horizontal coordinates $(x,y)$ for determining the center of the mask in the sky,   i.e. $x = a_0 + x' \cos\theta + y' \sin\theta$  \,and\, $y=b_0- x'\sin\theta+y'\cos\theta$.}

\subsection{Deciding the center, the rotation and the objects}
\label{se:angle}

\RefA{2}{The angle $\theta$ affects the position of the objects in the catalogue when rotating them to
work on an axis-parallel coordinate system. The rotated coordinates are determined by the system (\ref{eq:rot}).
Clearly, having $\theta$ unfixed implies having $(a'_i,b'_i)$ unknown values.
Introducing two continuous variables $s$ and $c$,
we can avoid having the sine and cosine functions, and work with the alternative system
\begin{eqnarray}
a'_i = (a_i-a_0) c -(b_i-b_0)s && \qquad \forall i\in I \label{eq:a1}\\
b'_i = (a_i-a_0) s +(b_i-b_0)c &&\qquad \forall i\in I \label{eq:a2}\\
s^2 + c^2 = 1 &&\label{eq:a3}\\
-1\leq s\leq 1&&\label{eq:a4}\\
-1\leq c\leq 1&&\label{eq:a5}
\end{eqnarray}
The lower bound in (\ref{eq:a4}) can be $0$ when the OSP in the spectrograph is vertically symmetric.}

\RefA{2}{The above consideration suggests the following  model for the MDP.
Consider the continuous variables $(s,c)$ representing the angle and $(x',y')$ determining the axis-parallel position
of the center of the mask respect to the centroid of the catalogue.
Consider also a pair of continuous variables $(a'_i,b'_i)$ for each $i\in I$ determining the axis-parallel position
of the object $i$ respect to the centroid of the catalogue, and
the binary variables $z_i$, $v_{ik}$, $w_{ik}$ and $u_{ij}$ as introduced in Subsection \ref{se:center}.
The objective function (\ref{eq:obj}) and the constraints (\ref{eq:x})--(\ref{eq:a5}) give a mathematical formulation for the MDP.
Recall that $(x',y')$ are rotated coordinates, and the non-rotated coordinates (in the horizontal coordinate system) for the center of the mask in the sky are $(x,y)$ with $x = a_0 + x' c + y' s$  \,and\, $y=b_0- x's+y'c$. The rotation angle $\theta$ is obtained from $c=\cos\theta$ and $s=\sin\theta$.}

This formulation for the MDP includes the non-convex equation (\ref{eq:a3}), leading to a non positive-defined matrix, which creates serious troubles for any general-purpose optimizer. Today there exist software with techniques to add and handle the sine and cosine of a variable angle $\theta$ (see e.g. \texttt{addGenConstrSin()}  and \texttt{addGenConstrCos()} in \emph{Gurobi}\footnote{\url{https://www.gurobi.com/}}, or \texttt{xpress.sin()} and \texttt{xpress.cos()} in \emph{Xpress}\footnote{\url{https://www.fico.com/}}), but our
preliminary computational experiments showed worse results than dealing with the variables $s,c$ and the quadratic equation (\ref{eq:a3}). Still, even this variant with $s$ and $c$ performed badly on instances of realistic size,
and it does not prevent off-source points too close to other objects in the sky because the angle is unfixed.
On the other hand, the number of angles to explore in the real-world application is quite small: many spectrographs only accept North-up masks (the parallactic angle) or very few options
(e.g., horizontal or perpendicular to the zenith).
As an example, the instructions for designing masks for the instrument OSIRIS include
``Although other angles are in principle possible, for the time being only slit orientations N-S or E-W will be accepted.''\footnote{\url{http://www.gtc.iac.es/instruments/osiris/osirisMOS.php}}
For those reasons,
\RefA{3}{we do not propose to solve the non-convex MDP model to optimality, but
to generate a heuristic MDP solution by enumerating a few values for $\theta$ and solving the COSP formulation
(\ref{eq:obj})--(\ref{eq:real}) for each of them. The proposal is described in the next section.}

\section{Algorithm}
\label{se:algorithm}


This section describes a two-phase heuristic approach for solving the MDP, which means finding $(\theta,x,y)$ such that the OSP with rotation angle $\theta$ and with the center pointing to $(x,y)$ gives the maximum profit (i.e. sum of priorities of the selected objects).
This section describes a heuristic approach assuming that $\theta$ is allowed to assume any value, but it can be adapted easily to find $(x,y)$ for a given $\theta$ if required.
The first phase generates an initial heuristic solution to be used in the second phase to solve the formulation (\ref{eq:obj})--(\ref{eq:real}). By doing so, the second phase may change the center $(x,y)$ but will not change the angle $\theta$ found in the first phase. Embedding the two-phase approach in a multi-start scheme could generate better solutions at the cost of increasing the computational time. Our experiments have shown us that even the first phase is enough to create excellent masks in our real-world application, so we did not have the need of the multi-start scheme.

We opted for designing an Iterative Local Search (ILS) approach to compute the initial heuristic solution in the first phase. ILS is a very simple and robust framework to explore the feasible domain of complex optimization problems,
\RefA{4}{and the literature contains many articles with successful implementations to a wide variety of problems; see e.g. \cite{ILS2018}.}
ILS has three main ingredients: a constructive procedure to start the iterative procedure with a promising solution, a local search procedure to investigate the local neighbourhood of a solution, and a perturbation mechanism to escape from a local optimum. The three ingredients may change $(\theta,x,y)$ and the associated OSP instance is solved to evaluate the corresponding solution profit. Each time the angle $\theta$ is changed, we rotate the coordinates of the catalogue using (\ref{eq:rot}), so the next OSP instances look for axis-parallel masks. Based on preliminary experiments on real-world data, and trying to balance exploration and computational time, we found convenient to change $\theta$ at high-level of a procedure and $(x,y)$ at low-level, thus trying to reduce the number of coordinate rotations.
\RefA{5}{We next describe each of the three ingredients of our ILS approach.}

The constructive procedure looks for a good solution by randomly exploring a grid of vertices in a cube.
One dimension of the cube represents the 180 degrees of a half-turn, where $\theta$ takes a value.
The other two dimensions conform the minimum-area rectangle where the object's coordinates $(a_i,b_i)$ are.
The grid is created by exploring the $\theta$ dimension on the multiples of $180/f$ degrees, and
the other two dimensions divided, each one, in $g+1$ identical intervals.
The parameter $f$ and $g$ are a-priori fixed (say $f=36$ and $g=1000$) and the grid contains $f\cdot g^2$ vertices,
which may be a very large number.
The constructive procedure randomly selects $p$ angles among the $f$ options, and for each angle $\theta$ it also randomly selects $q$ points $(x,y)$ to solve the OSP instance on each $(\theta,x,y)$.
The parameter $p$ and $q$ are also a-priori fixed (say $p=10$ and $q=100$).
The mask with the maximum profit among the $p\cdot q$ ones is selected as initial MDP solution.

The local search procedure explores the neighbourhood of the incumbent solution $(\theta,x,y)$ by slightly changing each component. At high-level we explore $f$ angles, at distance of one degree, around $\theta$. At low-level we consider the vertical and horizontal lines, and the two diagonals of the CSU, thus defining eight directions to move the current center $(x,y)$; the procedure considers each of these directions, computes the maximum $d$ that the center can be moved in that direction without losing a selected object, and solves the OSP with a new center obtained by the current one moved $l\cdot d$ arcseconds in the selected direction for each $l=1,\ldots,5$. Whenever a new mask is generated, the incumbent solution is updated and the procedure is applied again until the local search becomes trapped at a locally optimal solution.

The perturbation mechanism uses a random operator to change the incumbent solution $(\theta,x,y)$.
To this end, we define a perturbation strength $s$ to add to each component.
The value of $s$ is fundamental: a too small value will not help to escape from a locally optimal solution;
a too large value makes difficult to discover better quality solutions.
It must balance between intensification and diversification.
Based on preliminary experiments we found useful to change $\theta$ to $\theta+90$ and move $(x,y)$ to the symmetric position respect to the centroid $(a_0,b_0)$ of the catalogue. The components to change are randomly selected in each call to this perturbation mechanism.

While alternating between the perturbation mechanism and the local search procedure,
an acceptance criterion decides whether to continue or quit the iterative scheme.
Our acceptance criterion is based on the number of iterations without finding a mask with higher profit.

The best mask from the ILS approach is given as a starting primal solution to a general-purpose solver that solves the CSOP formulation (\ref{eq:obj})--(\ref{eq:real}) within a time limit in the second phase.
The ILS solution provides an initial lower bound on the optimal CSOP value. To further help the solver with tighter lower bounds, a primal heuristic approach is applied at each node of the branch-and-bound exploration. This approach consists of solving an OSP instance using the center of the optimal fractional solution of the LP relaxation in that branch-and-bound node.
We skip solving an OSP instance if the center of the fractional solution coincides with the center of the previously-solved OSP instance.


\section{Computational results}
\label{se:results}

We implemented  the algorithm in Python 3.7 using Gurobi 9.0 and a personal computer with Intel(R) Xeon(R) W-2125 CPU @ 4.00GHz, 64-bit Microsoft Windows 10 and Anaconda 3.
The performance was evaluated on two families of instances.

\begin{table}
\centering
\scriptsize
\begin{tabular}{ccccc}
name     & stars   & interest & brilliant & chop-and-nod \\ \hline
\texttt{l10}    & 32779   &  3220    & 1538      & 1682   \\
\texttt{l33}    & 30847   &  5144    & 2576      & 2568   \\
\texttt{l60}    & 14169   &  6127    & 891       & 5236   \\
\texttt{l10-15} & 54574   &  8017    & 2105      & 5912   \\ \hline
\end{tabular}
\caption{Characteristics of the read-world catalogues}\label{tab:catalogues}
\end{table}

The first family of instances is based on four real-world catalogues provided to us by IAC (Canary Islands, Spain) with the features in Table \ref{tab:catalogues}. The dataset with these instances is publicly available; see \cite{Mendeley}.
Column \emph{stars} shows the number of objects in each catalogue.
Column \emph{interest} is the number of objects of interest to the astronomer.
Column \emph{brilliant} is the number of objects of interest requiring the on-source observation only.
Column \emph{chop-and-nod} is the number of objects of interest also requiring the off-source observations.
This paper has a supplementary material containing these data files that, for each astronomical object,
include the identification label $i$, the 2-dimensional coordinates $(a_i,b_i)$ in arc-degrees, the priority $q_i$
(positive for objects needing the chopping-and-nodding technique, and negative for the others),
and the parameters $\underline W_i$ and $\overline W_i$ when a specific wavelength interval must be analyzed.
An object with priority 1 is preferred for the mask than another object with priority 2, and in general
an object $i$ is preferred than object $j$ if $|q_i|<|q_j|$.
We defined the profit $p_i = \max\{ |q_j|:i\in I \} - |q_i|$ for each object $i\in I$.
The recommended chopping-and-nodding throw for the non-brilliant objects is $t=4$ arcseconds.
The astronomical instrument is the EMIR spectrograph at the GTC telescope (Canary Islands, Spain).
Although its CSU is composed by 55 pairs of bars,
the technicians do not use the first and the last bands when the instrument operates in MOS mode.
The reason is that the projection of the spectrum of an object onto the detector is displaced
in the dispersion direction in accordance with the vertical position of the specific slitlet.
Thus, for slits close to the upper and lower boundaries of the field of view, part of the spectrum may be lost.
For that reason we were suggested to use $m=53$ on these catalogues.
The height of each band of EMIR is $h=7.235$ arcseconds,
the heights of the upper and lower zones in a band are $\underline h = \overline h = 1$ arcsecond,
and the width of a band is $W=240$ arcseconds.

The second family of instances consists of randomly-generated instances created with three numbers of bands, three numbers of objects of interest, and four values for the chopping-and-nodding throw. In particular, $m\in \{10,15,20\}$, $n\in \{500,1000,2000\}$, $t\in\{5,10,15,20\}$, and for each $(m,n,t)$ we have generated ten instances with different seeds for the generator of random numbers. The coordinates $(a_i,b_i)$ of the objects are random pairs of integers in $[-200,200]\times[-200,200]$ and the profits $p_i$ are random integers in $[5,10]$
\RefA{6}{inspired by real-world data (where the profit of an object is related to its position in a Colour Magnitude Diagram).}
Half of the objects are assumed to be brilliant (i.e. they do not need the chopping-and-nodding technique)
and the other half need the on- and off-source observations. We use $\underline W_i = -W/2$ and $\overline W_i = W/2$ for each object $i$ in this second family. The other parameters were set as in the first family of instances.

\begin{figure}[h]
\begin{subfigure}{.55\textwidth}
\centering\hspace{-1cm}
\includegraphics[width=\linewidth,keepaspectratio]{"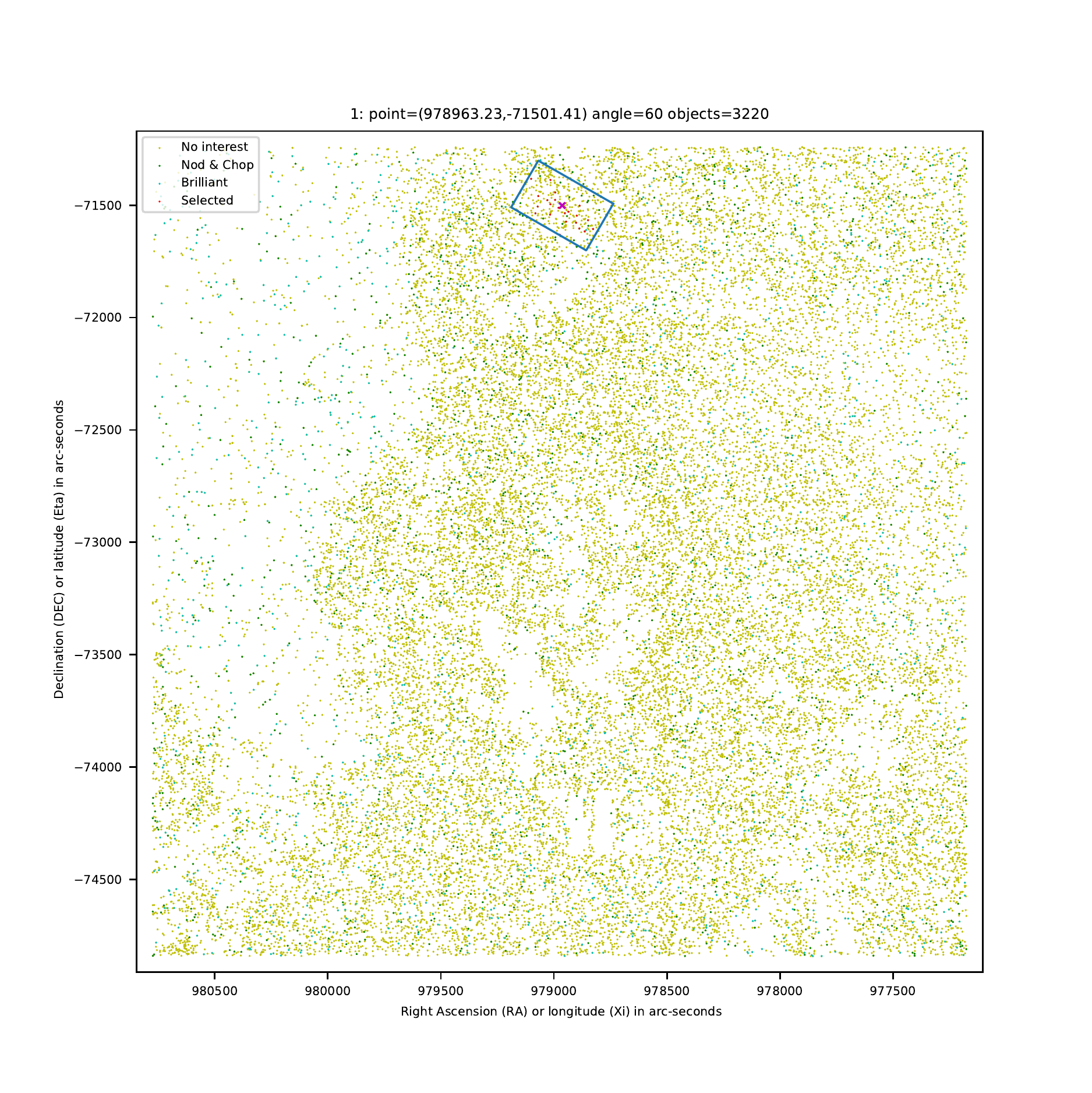"}
\caption{Field of view in the sky ($n=3220$)}
\label{fi:real10a}
\end{subfigure} \hspace{-1cm}
\begin{subfigure}{.55\textwidth}
\centering\vspace{-1.2cm}
\includegraphics[width=\linewidth,keepaspectratio]{"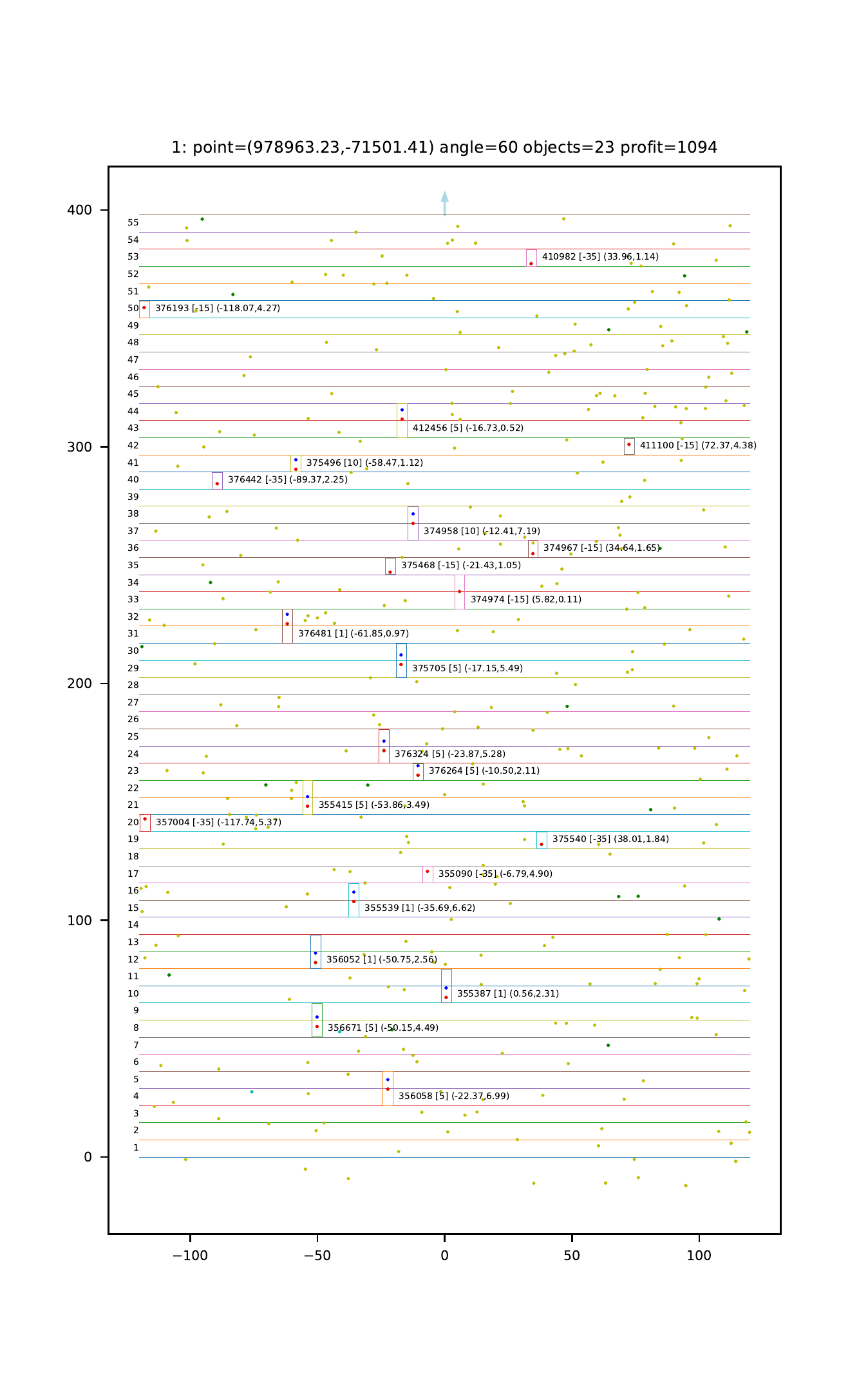"} \vspace{-2cm}
\caption{Mask}
\label{fi:real10b}
\end{subfigure}
\caption{Best solution for the real-world instance \texttt{l10}}
\label{fi:real10}
\end{figure}

\begin{figure}
\begin{subfigure}{.55\textwidth}
\centering\hspace{-1cm}
\includegraphics[width=\linewidth,keepaspectratio]{"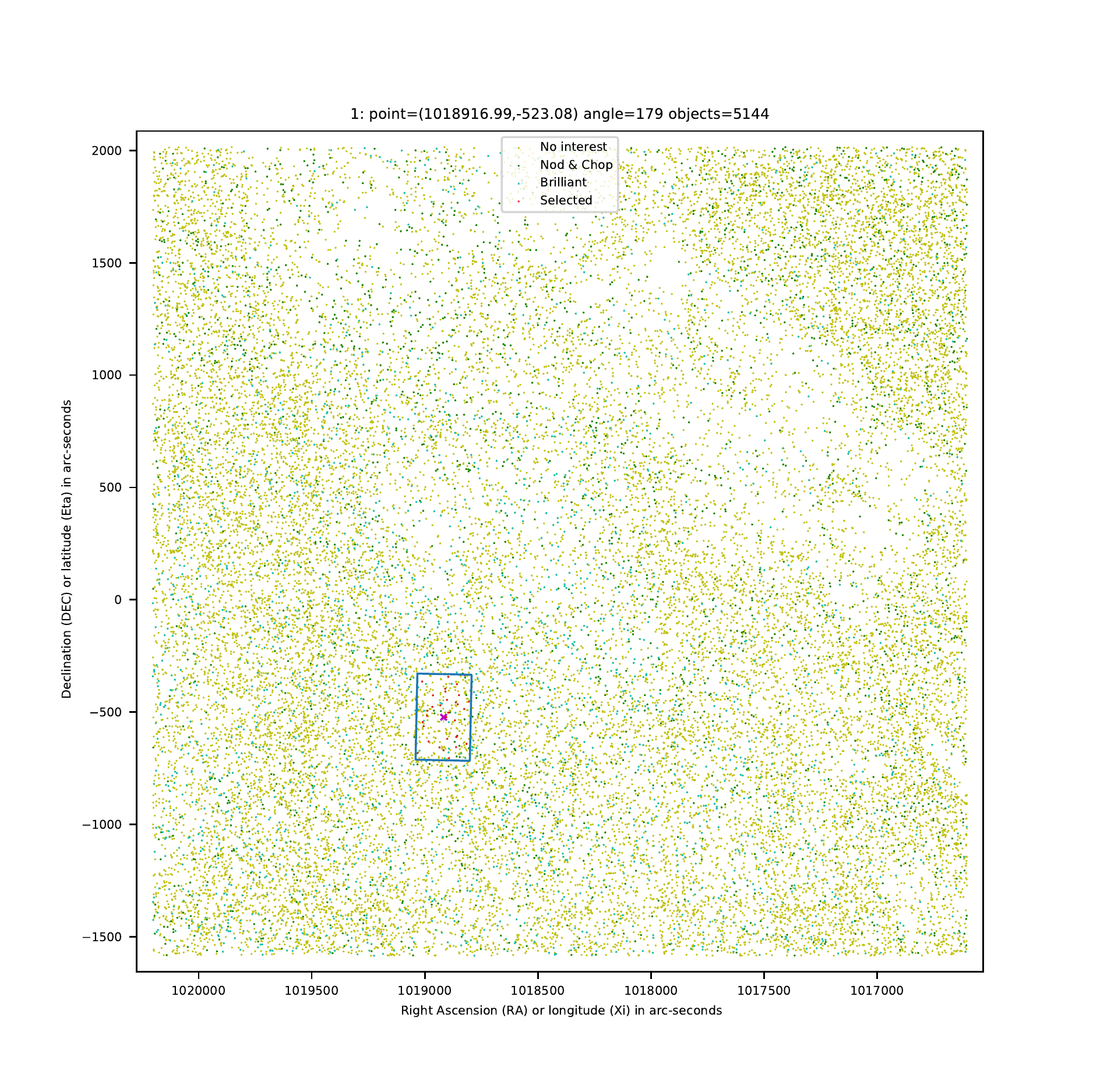"}
\caption{Field of view in the sky ($n=5144$)}
\label{fi:real33a}
\end{subfigure} \hspace{-1cm}
\begin{subfigure}{.55\textwidth}
\centering\vspace{-1.2cm}
\includegraphics[width=\linewidth,keepaspectratio]{"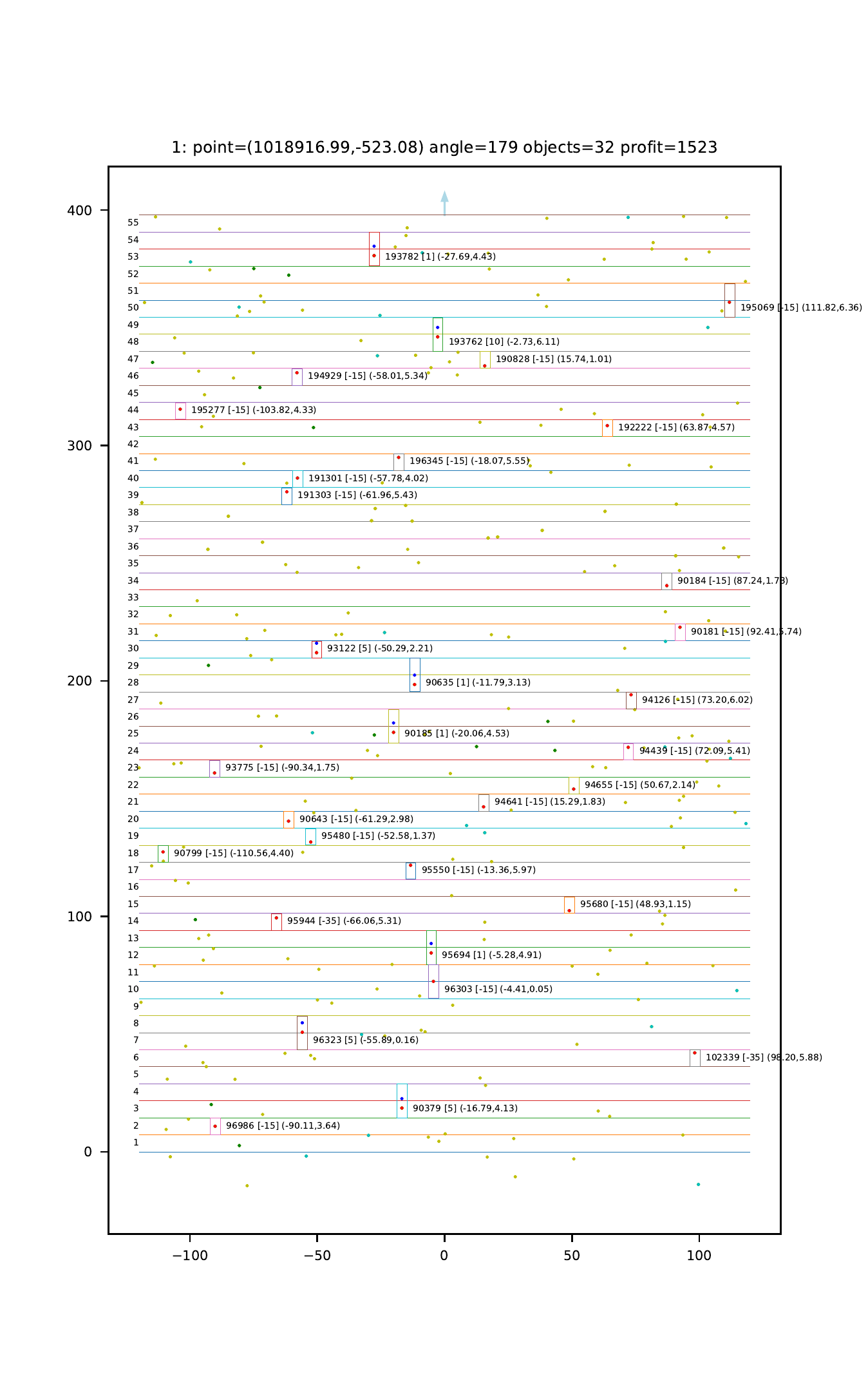"} \vspace{-2cm}
\caption{Mask}
\label{fi:real33b}
\end{subfigure}
\caption{Best solution for the real-world instance \texttt{l33}}
\label{fi:real33}
\end{figure}

\begin{figure}
\begin{subfigure}{.55\textwidth}
\centering\hspace{-1cm}
\includegraphics[width=\linewidth,keepaspectratio]{"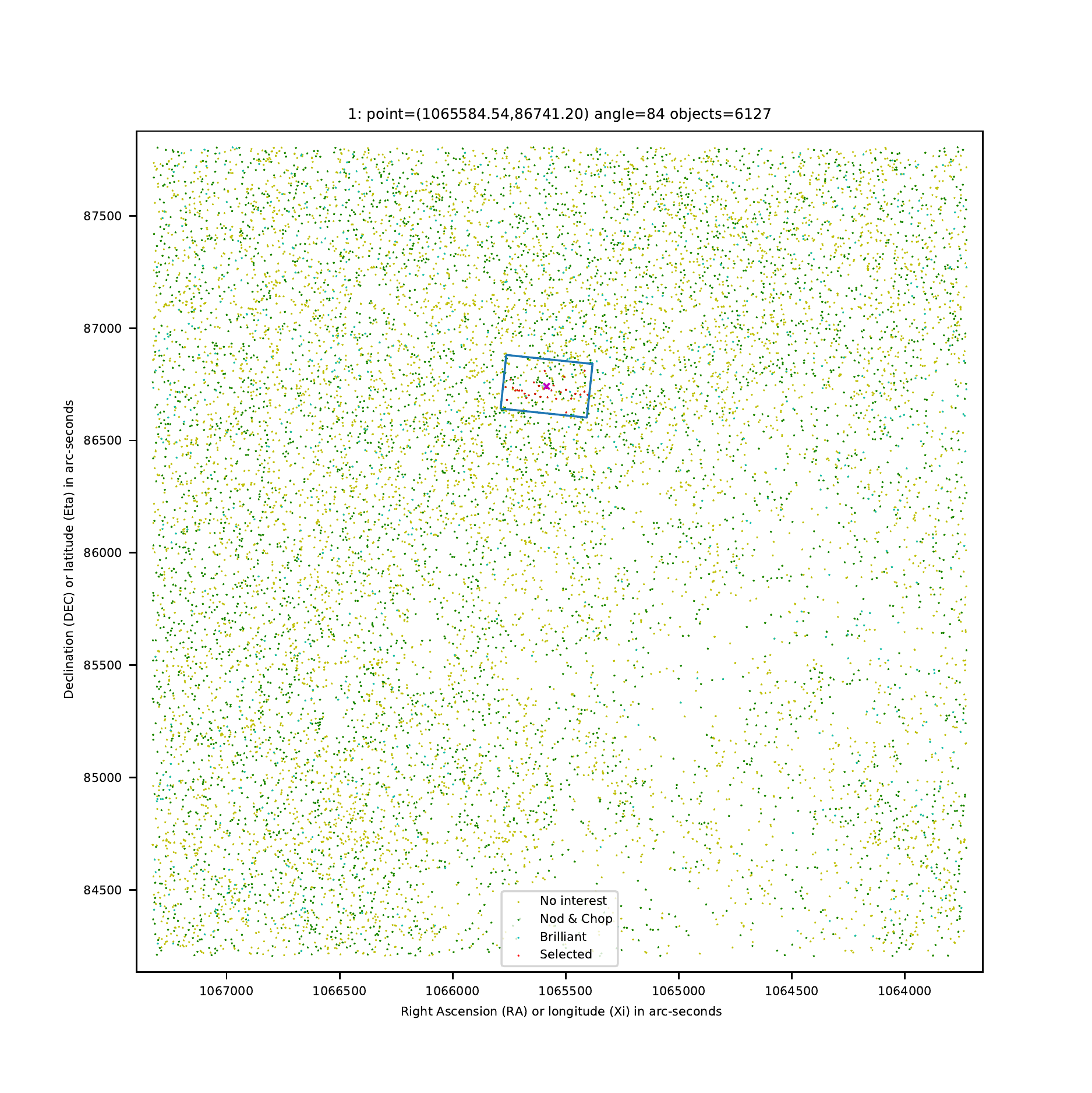"}
\caption{Field of view in the sky ($n=6127$)}
\label{fi:real60a}
\end{subfigure} \hspace{-1cm}
\begin{subfigure}{.55\textwidth}
\centering\vspace{-1.2cm}
\includegraphics[width=\linewidth,keepaspectratio]{"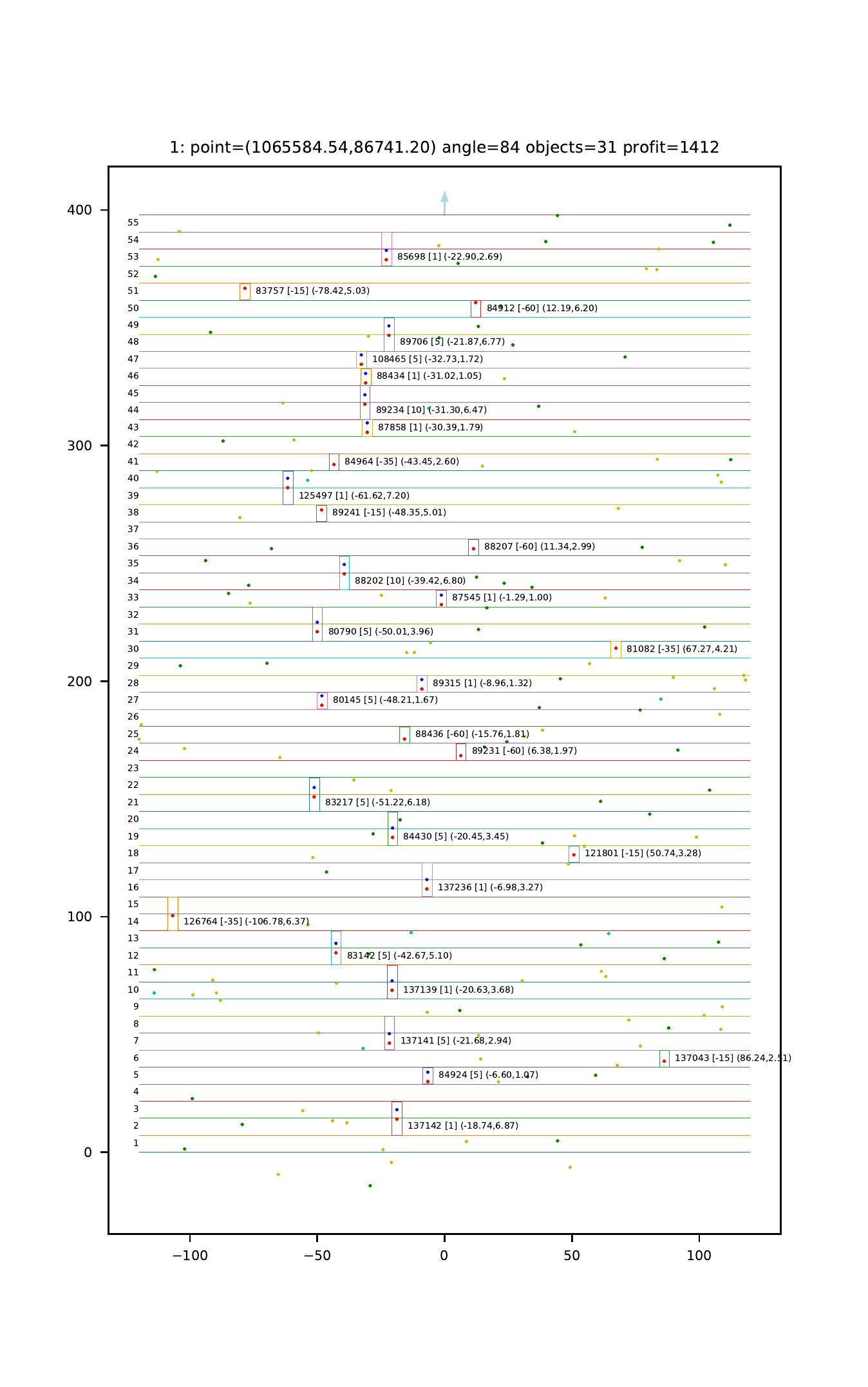"} \vspace{-2cm}
\caption{Mask}
\label{fi:real60b}
\end{subfigure}
\caption{Best solution for the real-world instance \texttt{l60}}
\label{fi:real60}
\end{figure}

\begin{figure}
\begin{subfigure}{.55\textwidth}
\centering\hspace{-1cm}
\includegraphics[width=\linewidth,keepaspectratio]{"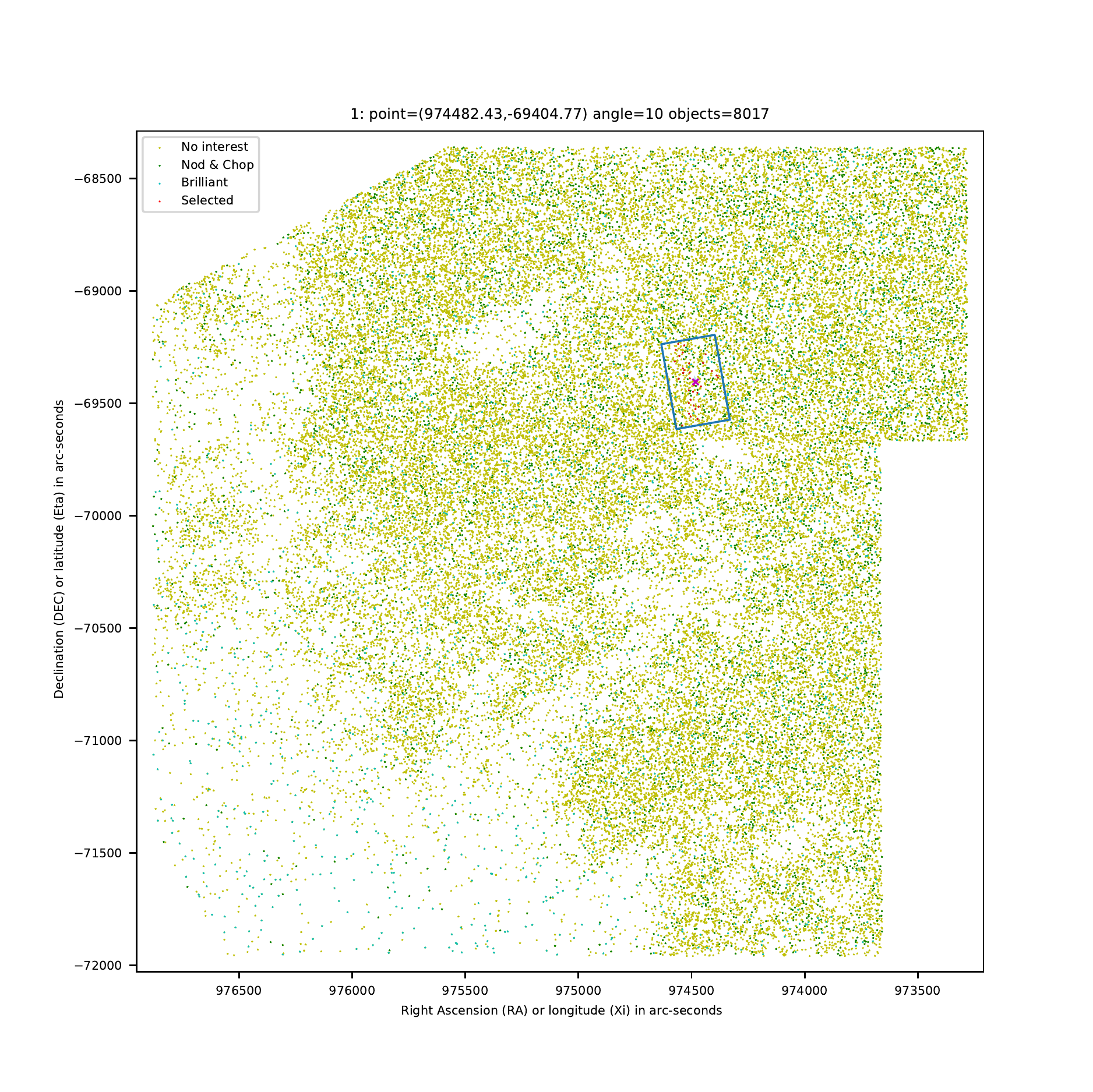"}
\caption{Field of view in the sky ($n=8017$)}
\label{fi:real10-15a}
\end{subfigure} \hspace{-1cm}
\begin{subfigure}{.55\textwidth}
\centering\vspace{-1.2cm}
\includegraphics[width=\linewidth,keepaspectratio]{"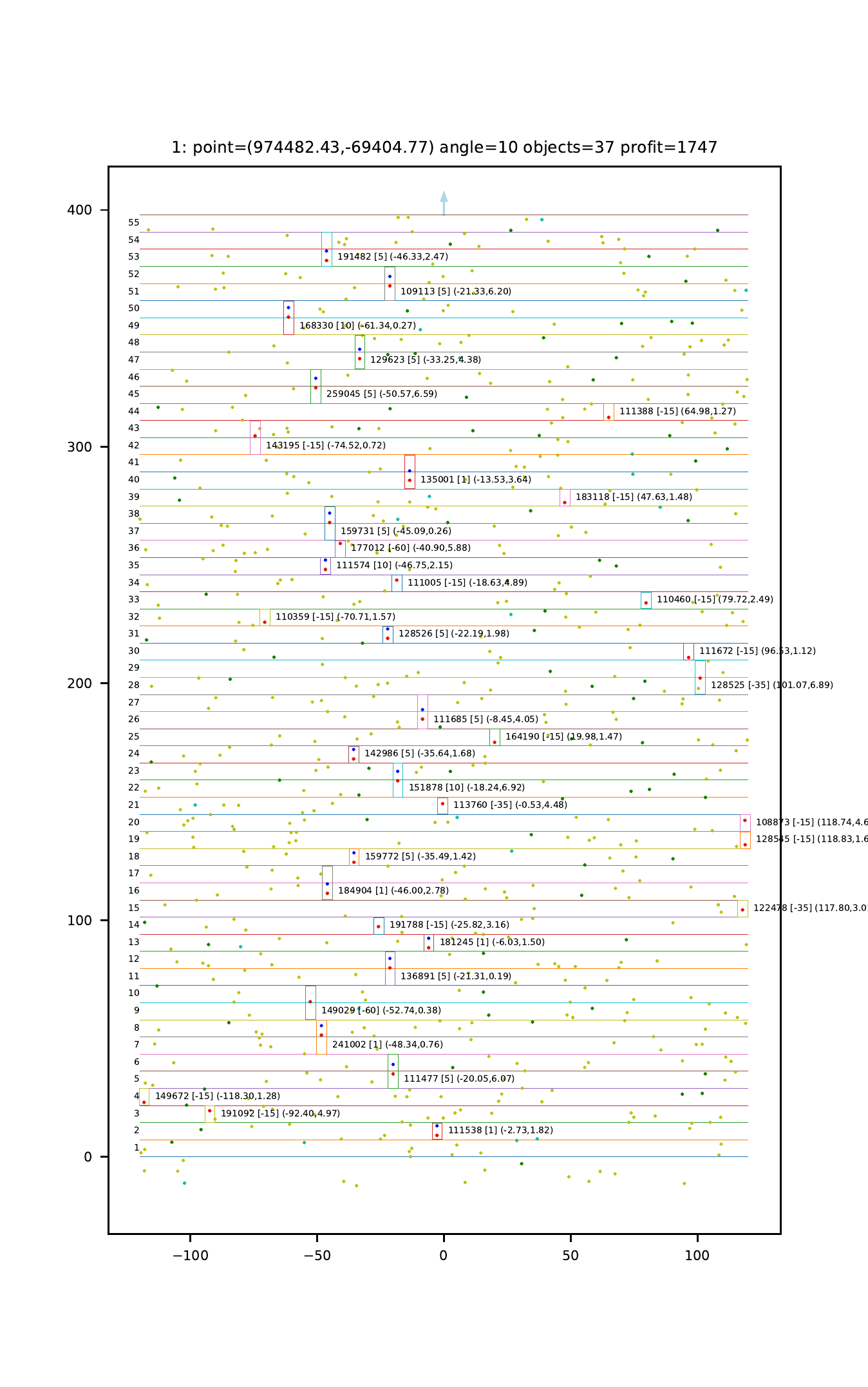"} \vspace{-2cm}
\caption{Mask}
\label{fi:real10-15b}
\end{subfigure}
\caption{Best solution for the real-world instance \texttt{l10-15}}
\label{fi:real10-15}
\end{figure}

\begin{table}
\centering
\scriptsize
\begin{tabular}{cc cc cc cc cc cc c}
  name   &  $|I|$   & $|I^1|$ &  $|I^2|$  &     sel1 &   sel2 & $x$          &   $y$         &   $\theta$   &    LB0   &  time0      &   vars   &    cons   \\
\hline
    \texttt{l10}  &  3220  &   1538  &   1682  &     10   &    13  &  978963.2306 &  -71501.4138  &    60.0000   &   1094  & 1702.9062   &   607076 &   1219107   \\
    \texttt{l10}  &  3197  &   1528  &   1669  &     20   &     7  &  979797.1072 &  -74122.5131  &   177.0000   &   1080  & 1408.5000   &   602584 &   1210087   \\
    \texttt{l10}  &  3170  &   1508  &   1662  &     21   &     6  &  977609.8973 &  -71511.7799  &    24.0000   &   1132  & 1365.9844   &   598616 &   1202117   \\
    \texttt{l10}  &  3143  &   1487  &   1656  &     22   &     5  &  979989.4146 &  -74602.8015  &    16.0090   &   1070  & 1362.4531   &   594806 &   1194464   \\
    \texttt{l10}  &  3116  &   1465  &   1651  &      5   &    15  &  979301.2182 &  -71532.9560  &    90.0000   &   1003  & 1369.0469   &   591154 &   1187128   \\ \hline
    \texttt{l33}  &  5144  &   2576  &   2568  &     24   &     8  & 1018916.9939 &    -523.0768  &   178.8153   &   1523  & 3083.1562   &   951008 &   1909781   \\
    \texttt{l33}  &  5112  &   2552  &   2560  &     28   &     9  & 1019819.5501 &    -956.1852  &    11.0795   &   1484  & 2906.8906   &   946352 &   1900429   \\
    \texttt{l33}  &  5075  &   2524  &   2551  &     23   &     9  & 1018154.5174 &    -373.2820  &    31.0000   &   1454  & 2500.3281   &   941008 &   1889695   \\
    \texttt{l33}  &  5043  &   2501  &   2542  &     25   &    11  & 1018475.7862 &    -984.0472  &    20.0352   &   1473  & 2203.4062   &   936194 &   1880026   \\
    \texttt{l33}  &  5007  &   2476  &   2531  &     27   &     6  & 1019277.0500 &     153.5776  &    88.0000   &   1436  & 2179.7656   &   930640 &   1868871   \\ \hline
    \texttt{l60}  &  6127  &    891  &   5236  &     11   &    20  & 1065584.5392 &   86741.1957  &    84.0029   &   1412  & 2238.8125   &  1476750 &   2964916   \\
    \texttt{l60}  &  6096  &    880  &   5216  &      8   &    20  & 1066397.7461 &   87583.7204  &   108.9907   &   1422  & 2284.7812   &  1470304 &   2951973   \\
    \texttt{l60}  &  6068  &    872  &   5196  &     10   &    18  & 1066562.2131 &   84356.9650  &    60.0000   &   1383  & 2163.2344   &  1464176 &   2939669   \\
    \texttt{l60}  &  6040  &    862  &   5178  &      8   &    18  & 1065306.5889 &   87305.1841  &   130.0000   &   1378  & 2110.0781   &  1458364 &   2927999   \\
    \texttt{l60}  &  6014  &    854  &   5160  &      9   &    18  & 1065396.9136 &   86969.8422  &    70.0000   &   1375  & 2072.2031   &  1452764 &   2916755   \\ \hline
 \texttt{l10-15}  &  8017  &   2105  &   5912  &     18   &    19  &  974482.4251 &  -69404.7741  &    10.0009   &   1747  & 4912.7969   &  1783898 &   3581778   \\
 \texttt{l10-15}  &  7980  &   2087  &   5893  &      5   &    26  &  974096.4147 &  -68593.2109  &    48.0205   &   1703  & 4896.5781   &  1776974 &   3567874   \\
 \texttt{l10-15}  &  7949  &   2082  &   5867  &     11   &    25  &  973648.3575 &  -69524.5958  &   107.0000   &   1829  & 4552.4844   &  1769580 &   3553029   \\
 \texttt{l10-15}  &  7913  &   2071  &   5842  &     21   &    16  &  974079.6385 &  -69068.8703  &    40.0000   &   1691  & 4137.1719   &  1761814 &   3537436   \\
 \texttt{l10-15}  &  7876  &   2050  &   5826  &     10   &    22  &  973778.4939 &  -70165.2609  &   158.0000   &   1700  & 4101.8125   &  1755364 &   3524483   \\ \hline
\end{tabular}
\caption{Real-world instances with $m=53$ and $t=4$}\label{tab:real}
\end{table}
Table \ref{tab:real} shows results of our implementation on five instances based on each catalogue of the first family.
For each catalogue, each instance was created by removing the selected objects in the previous instance, except the first one which deals with all the objects in the original catalogue.
In this way, the five instances from the same catalogue lead to five masks that could go to the spectrometer in a night for collecting research data for the time awarded astronomer.
Column \emph{name} identifies the original catalogue of the MDP instance.
Column $|I|$ is the number of objects of interest, and is the sum of the numbers in Columns $|I^1|$ and $|I^2|$ for each instance.
Columns \emph{sel1} and \emph{sel2} show the number of selected objects from $I^1$ and $I^2$, respectively.
Columns $x,y,\theta$ describe the center and rotation angle of the generated mask.
Column $LB0$ shows the associated profit,
and Column \emph{time0} is the computational time (in seconds) consumed by the ILS approach.
We also tried to solve the mathematical formulation (\ref{eq:obj})--(\ref{eq:real}) with the rotation of the best mask within one hour of time limit, but the size of the CSOP formulation did not allow neither to find better masks nor to prove the optimality of the one generated by the initial heuristic approach.
Columns \emph{vars} and \emph{cons} show the numbers of variables and constraints, respectively, of the COSP formulation.
It is worth mentioning that the large number of objects in the sky of these catalogues and the tight values
of $\overline W_i$ and $\underline W_i$ make these MDP instances quite difficult to solve.
Still, the masks generated from the ILS approach were evaluated as excellent by the astronomers.
Figures \ref{fi:real10}--\ref{fi:real10-15} show the first mask for each catalogue.
The success of the ILS approach is due to the simplicity of OSP, \RefB{18}{which was able to solve} about 720,000 instances with
different centers and angles for each MDP instance. We also believe that the computational time of this heuristic approach can be strongly reduced by rewriting the computer code in C rather than in Python.

\begin{table}[t]
\hspace{-1cm}
\scriptsize
\begin{tabular}{rrrrr rrrrr rrrrr rr}
$m$ &   $n$   &    $t$ &  vars    &    cons  &    sel1  &     sel2 &   LB0    &    LB    &      UB   &   gap    &   time0  &   time   &   nodes   &   cback  &  frac & \# \\
\hline
  10 &    500 &      5 &    17482 &    35732 &      9.1 &      0.9 &     95.7 &     96.7 &      96.7 &     0.00 &    19.67 &    66.74 &     945.5 &    285.6 &   0.0 &  10 \\
  10 &    500 &     10 &    17482 &    35732 &     10.0 &      0.0 &     94.6 &     95.6 &      95.6 &     0.00 &    24.26 &   103.31 &    1205.3 &    389.1 &  79.7 &  10 \\
  10 &    500 &     15 &    17482 &    35732 &     10.0 &      0.0 &     94.4 &     95.6 &      95.6 &     0.00 &    24.22 &    87.65 &    1536.7 &    415.3 &  58.7 &  10 \\
  10 &    500 &     20 &    17482 &    35732 &     10.0 &      0.0 &     94.5 &     95.6 &      95.6 &     0.00 &    23.32 &    68.58 &    1223.4 &    352.2 &  22.5 &  10 \\   \hline
  10 &   1000 &      5 &    34906 &    71336 &      9.3 &      0.7 &     98.8 &     99.7 &      99.7 &     0.00 &    31.58 &    46.44 &     113.6 &     35.9 &   0.0 &  10 \\
  10 &   1000 &     10 &    34906 &    71336 &     10.0 &      0.0 &     98.8 &     99.3 &      99.3 &     0.00 &    37.52 &    31.38 &       0.7 &      2.3 &  16.0 &  10 \\
  10 &   1000 &     15 &    34906 &    71336 &     10.0 &      0.0 &     98.7 &     99.3 &      99.3 &     0.00 &    36.32 &    39.13 &       0.8 &      2.5 &  12.3 &  10 \\
  10 &   1000 &     20 &    34906 &    71336 &     10.0 &      0.0 &     98.4 &     99.3 &      99.3 &     0.00 &    35.16 &    40.90 &       4.8 &      6.8 &   0.9 &  10 \\   \hline
  10 &   2000 &      5 &    69812 &   142662 &      9.5 &      0.5 &    100.0 &    100.0 &     100.0 &     0.00 &    53.92 &    25.25 &       0.0 &      0.0 &   0.0 &  10 \\
  10 &   2000 &     10 &    69812 &   142662 &     10.0 &      0.0 &    100.0 &    100.0 &     100.0 &     0.00 &    61.55 &    40.46 &       0.0 &      0.0 &   0.0 &  10 \\
  10 &   2000 &     15 &    69812 &   142662 &     10.0 &      0.0 &    100.0 &    100.0 &     100.0 &     0.00 &    58.99 &    49.79 &       0.0 &      0.0 &   0.0 &  10 \\
  10 &   2000 &     20 &    69812 &   142662 &     10.0 &      0.0 &    100.0 &    100.0 &     100.0 &     0.00 &    56.09 &    43.52 &       0.0 &      0.0 &   0.0 &  10 \\   \hline
  15 &    500 &      5 &    26352 &    53477 &     13.9 &      1.1 &    138.8 &    140.1 &     147.5 &     5.34 &    26.11 &   399.68 &    4181.2 &   1165.4 &   0.0 &   3 \\
  15 &    500 &     10 &    26352 &    53477 &     15.0 &      0.0 &    136.9 &    137.6 &     148.4 &     7.92 &    30.35 &   290.94 &    2803.2 &    910.6 & 114.7 &   2 \\
  15 &    500 &     15 &    26352 &    53477 &     14.9 &      0.0 &    136.6 &    138.0 &     147.5 &     6.98 &    30.14 &   493.38 &    4308.3 &   1377.5 & 123.4 &   3 \\
  15 &    500 &     20 &    26352 &    53477 &     14.9 &      0.0 &    136.6 &    137.8 &     147.5 &     7.13 &    29.79 &   294.01 &    2981.0 &   1021.8 &  34.6 &   3 \\   \hline
  15 &   1000 &      5 &    52616 &   106761 &     13.6 &      1.4 &    146.4 &    147.7 &     148.5 &     0.55 &    41.67 &   404.50 &    3589.9 &   1123.4 &   0.0 &   8 \\
  15 &   1000 &     10 &    52616 &   106761 &     15.0 &      0.0 &    146.4 &    147.1 &     147.6 &     0.34 &    47.03 &   369.29 &    3043.8 &    980.6 &  20.2 &   9 \\
  15 &   1000 &     15 &    52616 &   106761 &     15.0 &      0.0 &    146.3 &    147.1 &     147.6 &     0.34 &    45.89 &   321.44 &    2613.9 &    909.4 &  22.9 &   9 \\
  15 &   1000 &     20 &    52616 &   106761 &     15.0 &      0.0 &    146.1 &    147.1 &     147.6 &     0.34 &    45.35 &   340.40 &    2648.7 &    885.4 &   1.0 &   9 \\   \hline
  15 &   2000 &      5 &   105232 &   213507 &     14.7 &      0.3 &    149.8 &    150.0 &     150.0 &     0.00 &    71.09 &    77.83 &       0.2 &      0.4 &   0.0 &  10 \\
  15 &   2000 &     10 &   105232 &   213507 &     15.0 &      0.0 &    149.8 &    150.0 &     150.0 &     0.00 &    78.70 &    91.15 &       0.2 &      0.2 &   0.0 &  10 \\
  15 &   2000 &     15 &   105232 &   213507 &     15.0 &      0.0 &    149.7 &    150.0 &     150.0 &     0.00 &    77.25 &   122.67 &       0.3 &      0.4 &   0.0 &  10 \\
  15 &   2000 &     20 &   105232 &   213507 &     15.0 &      0.0 &    149.9 &    150.0 &     150.0 &     0.00 &    76.43 &   115.11 &       0.1 &      0.0 &   0.0 &  10 \\   \hline
  20 &    500 &      5 &    35222 &    71222 &     18.3 &      1.6 &    178.8 &    180.9 &     200.0 &    10.60 &    31.08 &     --   &    2791.1 &    940.8 &   0.0 &   0 \\
  20 &    500 &     10 &    35222 &    71222 &     19.2 &      0.3 &    177.1 &    178.2 &     200.0 &    12.32 &    37.24 &     --   &    2326.3 &    798.8 & 143.7 &   0 \\
  20 &    500 &     15 &    35222 &    71222 &     19.3 &      0.2 &    176.8 &    177.6 &     200.0 &    12.70 &    37.84 &     --   &    2234.1 &    693.7 & 170.7 &   0 \\
  20 &    500 &     20 &    35222 &    71222 &     19.5 &      0.1 &    177.0 &    177.8 &     200.0 &    12.57 &    37.75 &     --   &    2178.1 &    775.5 &  44.7 &   0 \\   \hline
  20 &   1000 &      5 &    70326 &   142186 &     18.4 &      1.6 &    193.5 &    194.9 &     199.3 &     2.28 &    53.66 &   348.42 &    3157.8 &    982.5 &   0.0 &   3 \\
  20 &   1000 &     10 &    70326 &   142186 &     20.0 &      0.0 &    193.4 &    193.8 &     199.1 &     2.76 &    59.89 &   417.56 &    7638.5 &   2100.3 &  20.1 &   3 \\
  20 &   1000 &     15 &    70326 &   142186 &     20.0 &      0.0 &    192.9 &    193.9 &     199.1 &     2.70 &    60.88 &   527.37 &   10351.6 &   2887.8 &  26.5 &   3 \\
  20 &   1000 &     20 &    70326 &   142186 &     20.0 &      0.0 &    193.3 &    193.9 &     199.1 &     2.71 &    59.18 &   437.75 &    6650.9 &   2002.9 &   0.7 &   3 \\   \hline
  20 &   2000 &      5 &   140652 &   284352 &     18.9 &      1.1 &    199.4 &    199.8 &     199.8 &     0.00 &    94.80 &   136.72 &       0.5 &      1.4 &   0.0 &  10 \\
  20 &   2000 &     10 &   140652 &   284352 &     20.0 &      0.0 &    199.2 &    199.5 &     199.5 &     0.00 &   108.71 &   192.43 &       1.8 &      3.5 &   0.0 &  10 \\
  20 &   2000 &     15 &   140652 &   284352 &     20.0 &      0.0 &    199.2 &    199.5 &     199.5 &     0.00 &   108.01 &   211.64 &       0.6 &      1.3 &   0.0 &  10 \\
  20 &   2000 &     20 &   140652 &   284352 &     20.0 &      0.0 &    199.3 &    199.5 &     199.5 &     0.00 &   115.58 &   238.47 &       0.6 &      1.2 &   0.0 &  10 \\   \hline

\end{tabular}
\caption{Random instances}\label{tab:random}
\end{table}

We conclude the computational analysis with the randomly-generated instances.
Table \ref{tab:random} summarizes the results of executing our computer code on our second family of instances.
The aim of these experiments was to measure the performance of the exact approach, based on the mathematical formulation
(\ref{eq:obj})--(\ref{eq:real}), and also the quality of the ILS approach on instances which we can solve to optimality. To that end we fixed $\theta=0$, hence only one CSOP formulation is solved for each instance after having applied the initial heuristic approach.
Each line of the table shows the average results of the ten instances for each $(m,n,t)$.
Some columns are the ones described before, although now they show average values. In addition, the table also shows
Columns $UB$ and $LB$, which are the best upper and lower bounds after having solved formulation (\ref{eq:obj})--(\ref{eq:real}).
Column \emph{gap} is the percentage deviation between $LB$ and $UB$, with zero meaning that the generated mask is optimal.
Column \emph{nodes} is the number of explored branch-and-bound nodes by the solver, in addition to the root node,
and Column \emph{cback} is the number of OSP instances solved during the branch-and-bound exploration.
The number in \emph{cback} is smaller than the number in \emph{nodes} because some nodes
are fathomed internally by the solver, but also because we skip solving the OSP for producing a primal heuristic solution if the center
of the current fractional solution coincides with the center of the previous OSP solved.
About half of the computational time was consumed by our procedure for solving the OSP instances.
We imposed a time limit of 10 minutes to the solver on each COSP formulation, and Column \# shows the number over ten that were solved without achieving this limit. Column \emph{time} was computed as the average over
the ones solved before the time limit, and it shows ``$-$'' when no one finished with the optimality proof of the final mask.
Column \emph{frac} is the average number of solved OSP instances (while computing the initial ILS mask and
while exploring the branch-and-bound tree) with a non-integer optimal solution when only the LP relaxation of (\ref{eq:w1})--(\ref{eq:w3}) is solved.
Thus, Column \emph{frac} shows the average number of OSP instances on which the blossom inequalities (\ref{eq:blossom}) are used.

Table \ref{tab:random} shows that the major impact on the performances of the algorithm is due to
the number of bands $m$.
Taking into account that the average number of solved OSP instances is quite large (about 20,000 instances in the heuristic approach, plus the number in \emph{cback}),
the small numbers in Column \emph{frac} show that the LP relaxation of (\ref{eq:w1})--(\ref{eq:w3}) gives an integer OSP solution in most cases.
While all CSOP instances are solved to optimality when $m=10$, the execution goes to the
time limit on around one-third of the instances with \RefB{20}{$m=15$,} and on around two-thirds of the instances with \RefB{20}{$m=20$.}
The impact of the number of objects on the heuristic approach and on the formulation is quite different.
In a sense, it is easier to find feasible solutions when there are more objects, although there is a cost
for managing the objects inside the computer code. When solving the mathematical formulation, a larger instance
needs more variables and more constraints in the formulation, but the LP relaxation is better.
Indeed, all instances with $n=2000$ were solved to optimality.
\RefB{19}{A potential explanation for this behaviour is that a larger number of objects helps finding a good object for each band,
and indeed good heuristic solutions are earlier generated and the lower bounds from the linear programs are tighter.}
The computer code on these instances does not seem sensitive to the value of $t$,
in part because half of the catalogue consists of objects not needing the chopping-and-nodding technique.
Indeed, the selected objects for the optimal masks typically need the on-source observation only,
requiring the pair of bars of one band, and with $p_i=10$; for that reason the values in \emph{sel1} are close to $m$,
and the values in \emph{LB} and \emph{UB} are close to $10m$.
It is interesting to observe that the primal heuristic approach applied at each branch-and-bound node slightly
helps to improve the lower bound (from the values in \emph{LB0} to the values in \emph{LB}).
Considering the small time consumed to compute the initial mask (see Column \emph{time0})
and the proximity of $LB0$ to $LB$, the described ILS approach has reveled as an excellent algorithm for
solving the MDP.


\section{Conclusion}

The paper has addressed a new optimization problem motivated by a real-world application in astronomy.
The problem consists of deciding the translation and rotation of the field of view for an spectrograph with a configurable slit unit.
It also concerns deciding the subset of objects to observe. The aim is to maximize the profit of the selected objects.
The problem contains many requirements due to different types of objects in the catalogue,
some of them needing on- and off-source observations, and to the division of the field of view into bands, each one able to analyze the light of one object.
A very relaxed version of this problem is the location of a unit square to cover the maximum number of points of a given set, which is polynomial-time solvable.
When the translation and rotation are given, and only the object selection is unknown,
the problem is formulated as a maximum-weight matching problem.
When only the rotation angle is given, the problem is modelled with a Mixed Integer Linear Programming formulation.
We have described an ILS approach to solve the problem when the rotation is unfixed.
Computational experiments on random and real-world instances show that the approach works quite satisfactorily in practice.
The parameter with the highest impact on the performance is the number of bands of the spectrograph.
Optimal solutions can be obtained in short computational times when this number is up to 20 bands,
while in practice spectrometers have about 50 bands and then the ILS approach generates masks of high quality.
The computational analysis reveals that the weak element to prove the
optimality of a solution for the problem with a given rotation is the bound from the LP relaxation,
mainly because our formulation is based on big-$M$ values.
An interesting topic for future research is the need for alternative formulations with better relaxations.
\RefB{7}{Since the number of objects in a catalogue is finite, the number of values for $(\theta,x,y)$ is also finite,
thus a discretisation approach could deserve future investigations.}
Formulations in convex optimization with unfixed rotation would also be of interest.
\RefB{16}{Although the approach in this paper generates a single mask with maximum profit,
a more challenging problem is to generate a sequence of masks from a catalogue that can be used within the
awarded time of the telescope to the astronomer.}
While waiting for these alternatives, the algorithm described in this paper
is generating satisfactory masks in the real-world application.

\section*{Acknowledgement}
The author thanks Prof. Francisco Garzón López at the ``Instituto de Astrofísica de Canarias'' for posing and describing the topic in this paper, providing also with real data to evaluate the proposed algorithm.
This research was partially supported by the Spanish projects MTM2015-63680-R (MINECO/FEDER, UE), PID2019-104928RB-I00 (MINECO/FEDER, UE) and ProID2017010132 (Gobierno de Canarias/FEDER, UE).


\vspace{6cm}

\end{document}